%% file: main.tex
\documentclass[11pt,letterpaper]{article}
\usepackage[utf8]{inputenc}
\usepackage[T1]{fontenc}
\usepackage{amsmath}
\usepackage{amsfonts}
\usepackage{amssymb}
\usepackage{graphicx}
\usepackage[margin=1in]{geometry}
\graphicspath{{img/}}
\usepackage{booktabs}
\usepackage{hyperref}
\usepackage{breakcites}
\usepackage{caption}
\usepackage{subcaption}
\usepackage{multicol}
\usepackage{authblk}
\usepackage{bbm}
\usepackage{soulutf8}
\DeclareMathOperator*{\argmin}{arg\,min}

\input{tikz-options.tex}
\usepackage{enumitem}
\setlist{nosep}
\usepackage{tabularx}
\newcolumntype{s}{>{\hsize=.5\hsize}X}

\title{Rethinking Representations in P\&C Actuarial Science with Deep Neural Networks}

\author[1,3]{Christopher Blier-Wong}
\author[2,3]{Jean-Thomas Baillargeon}
\author[1,3,4]{Hélène Cossette}
\author[2,3]{Luc Lamontagne}
\author[1,3,4]{Etienne Marceau\thanks{Corresponding author: Etienne Marceau, etienne.marceau@act.ulaval.ca}}

\affil[1]{École d'actuariat, Université Laval, Québec, Canada}
\affil[2]{Département d'informatique et de génie logiciel, Université Laval, Québec, Canada}
\affil[3]{Centre de recherche en données massives, Université Laval, Québec, Canada}
\affil[4]{Centre interdisciplinaire en modelisation mathématique, Université Laval, Québec, Canada}

\date{\today}

\begin{document}
	
	\maketitle
	
	\input{abstract}
	\input{introduction}
	\input{data}
\input{representations}

	\input{constructing}
	\input{textual}
	\input{visual}
	\input{spatial}
	\input{conclusion}
	
	\bibliographystyle{abbrv}
	\bibliography{ref}


\end{document}

%% file: tikz-options.tex
\usepackage{tikz}

\usetikzlibrary{shapes.symbols, shapes.geometric, arrows, calc, backgrounds, shadows, trees, shapes, automata, positioning}
\usetikzlibrary{patterns}
\tikzset{
	basic/.style  = {draw,  drop shadow, rectangle, align=flush center},
	root/.style   = {basic, rounded corners=2pt, thin, align=flush center, fill=blue!30, text width=13em},
	level 2/.style = {basic, rounded corners=8pt, align=flush center, fill=blue!20, text width=8em, minimum height=30pt},
	level 3/.style = {basic,  align=flush left, fill=orange!10, text width=9em, rounded corners = 4pt}
}

\tikzset{%
	every neuronr/.style={
		circle,
		draw,
		fill = red!25,
		minimum size=0.5cm
	},
	every neuron/.style={
		circle,
		draw,
		minimum size=0.5cm
	},
	every neuronb/.style={
		circle,
		draw,
		fill = blue!25,
		minimum size=0.5cm
	},
	every neuronv/.style={
		circle,
		draw,
		fill = violet!50,
		minimum size=0.5cm
	},
	neuron missing/.style={
		draw=none, 
		scale=1,
		fill=white,
		text height=.333cm,
		execute at begin node=\color{black}$\vdots$
	},
}

\tikzstyle{data} = [rectangle, rounded corners, minimum width=4.5cm, minimum height=1cm,text centered, text width = 4.5cm, draw=black, fill=orange!30]
\tikzstyle{io} = [trapezium, trapezium left angle=70, trapezium right angle=110, minimum width=3cm, minimum height=1cm, text centered, draw=black, fill=violet!30]
\tikzstyle{process} = [rectangle, minimum width=4.5cm, minimum height=1cm, text centered, draw=black, text width = 4.5cm, fill=violet!30]
\tikzstyle{decision} = [diamond, minimum width=4.5cm, minimum height=3cm, text centered, text width=4.5cm, draw=black, fill=green!30]
\tikzstyle{decisiont} = [diamond, minimum width=3cm, minimum height=3cm, text centered, text width=3cm, draw=black, fill=orange!30]
\tikzstyle{arrow} = [thick,->,>=stealth]
\tikzstyle{datae} = [rectangle, rounded corners, minimum width=4.5cm, minimum height=1cm,text centered, text width = 4.5cm, draw=black, fill=green!30]
\tikzstyle{processt} = [rectangle, minimum width=4.5cm, minimum height=1cm, text centered, draw=black, text width = 4.5cm, fill=orange!30]

\pgfdeclarelayer{background}
\pgfdeclarelayer{foreground}
\pgfsetlayers{background,main,foreground}
\tikzstyle{arrow} = [thick,->,>=stealth]

\newcommand{\tstar}[5]{
	\pgfmathsetmacro{\starangle}{360/#3}
	\draw[#5, color = red] (#4:#1)
	\foreach \x in {1,...,#3}
	{ -- (#4+\x*\starangle-\starangle/2:#2) -- (#4+\x*\starangle:#1)
	}
	-- cycle;
}

\usetikzlibrary{fit,positioning,arrows.meta}
\tikzset{rnnneuron/.style={shape=circle, minimum size=1.25cm, 
		inner sep=0, draw, font=\small}, rnnio/.style={rnnneuron, fill=gray!20}}

\newcommand{\rnnvect}[2]{\boldsymbol{#1}^{\langle #2 \rangle}}
\newcommand{\parvect}[2]{\boldsymbol{#1}^{[ #2 ]}}

\usetikzlibrary{decorations.pathreplacing,3d,decorations.text,shapes.arrows,positioning,fit,backgrounds,chains,arrows.meta, shapes}

\usepackage{import}
\subimport{layers/}{init}

\usetikzlibrary{positioning}
\usetikzlibrary{3d} 

\makeatletter
\DeclareRobustCommand{\rvdots}{%
	\vbox{
		\baselineskip4\p@\lineskiplimit\z@
		\kern-\p@
		\hbox{.}\hbox{.}\hbox{.}
}}
\makeatother

%% file: layers/init.tex
\usetikzlibrary{quotes,arrows.meta}
\usetikzlibrary{positioning}

\def\edgecolor{rgb:blue,4;red,1;green,4;black,3}

\usepackage{Ball}
\usepackage{Box}
\usepackage{RightBandedBox}

%% file: abstract.tex
\begin{abstract}
	Insurance companies gather a growing variety of data for use in the insurance process, but most traditional ratemaking models are not designed to support them. In particular, many emerging data sources (text, images, sensors) may complement traditional data to provide better insights to predict the future losses in an insurance contract. This paper presents some of these emerging data sources and presents a unified framework for actuaries to incorporate these in existing ratemaking models. Our approach stems from representation learning, whose goal is to create representations of raw data. A useful representation will transform the original data into a dense vector space where the ultimate predictive task is simpler to model. Our paper presents methods to transform non-vectorial data into vectorial representations and provides examples for actuarial science. 	
\end{abstract}

\textbf{Keywords:} Feature learning, emerging data, unstructured data, embeddings, neural networks, machine learning, property and casualty insurance, pricing

%% file: introduction.tex
\section{Introduction}

Actuaries play essential roles in Property and Casualty (P\&C) insurance companies. From predicting future claims to managing enterprise risk, they combine their domain expertise with statistical models to achieve the company's objectives. At the center of these models are data: constructing predictive models requires historical experience, see Figure 8.1 in \cite{parodi2014pricing}.

This paper is about data. This paper is not about modeling, but rather about the crucial \textit{matière première} that is data. Insurance companies acknowledge that they sit on a precious \textit{natural resource}: data. Data is everywhere and has a significant impact on financial institutions. Abundant collections of data may improve the understanding of the composition of risks in their portfolio and provide more accurate predictions of losses (pricing), claim development (reserving) and dependence (risk management). Specifically, emerging data sources will reduce model heterogeneity by segmenting customers into more homogeneous risk classes. New external data sources also provide more accurate insurance premiums for new contracts in business segments with lower exposures. This emerging data source is sometimes called \textit{Big data} and consists of structured, unstructured and semi-structured data. Most existing ratemaking models are designed to use vectorial variables, a type of structured data that spreadsheets can neatly store. Actuaries expect underwriters to ask questions upon quoting the contract, and each answer has a dedicated column in the database, making statistical modeling with generalized linear models (GLMs) straightforward. Non-vectorial data, such as text, voice, images, sensors and other big data, require more processing before being used in such a statistical model. A recent series on predictive models in actuarial science \cite{frees2014predictive, frees2014predictive2} presents one application of unstructured data for pricing (a model for pricing telematics driving), indicating a potential lack of a framework for dealing with unstructured data for P\&C ratemaking.

This paper provides a framework to use non-vectorial data within traditional ratemaking models using representation learning embeddings. The representation of an observation corresponds to the information that is used to produce a prediction. In classical ratemaking, the representation of an \textit{a priori} loss estimate is the input variables provided by the potential customer during the quoting process. As companies collect and organize emerging data, these are added to existing representations, making the combined representations non-vectorial and unsuitable for existing ratemaking models. Representations operate on a more recent scientific paradigm where quantitative methods and massive emerging data can produce better predictions with a simpler model. The approach we take in this paper is to create vectorial representations from emerging data. 

It is worthwhile to differentiate between variables and features, which have different meanings in this paper. By variables, we mean the raw information collected from the insured. The amount and type of rating variables depend on the line of business, the insurer history and the actuary or underwriter's preference. By features, we mean the information that is used as inputs to the ratemaking model. A feature could be a variable itself but could also include modified or transformed versions of existing variables. For instance, the underwriter may ask the insured's age (or date of birth), which would be a variable. A feature typically used in a ratemaking model is the insured's age, so in this case, the variable is also a feature. An actuary could also consider other features in the model, including polynomial transformations of the age (like age squared) or the interaction between age and gender. Another example includes the marital status of the insured. The variable would be the qualitative (nominal) variable, while the feature would be a quantitative transformation (like the one-hot encoding of the class, see subsection \ref{ss:qualitative} for details).

\subsection{Objective}

The typical approach in ratemaking is to use statistical models with variables as inputs. We call this approach variable-based learning since the inputs to the models are the unmodified variables. Consider a dataset of $n$ observations each with a vector of variables $\boldsymbol{x}_{i}$ for $i = 1, \dots, n$. If the variables are vectorial, then the length of the vector $\boldsymbol{x}_{i}$ will be the same for all observations $i = 1, \dots, n$. For instance, if $p$ variables are available for each observation, then $\boldsymbol{x}_i = (1, x_{i1}, x_{i2}, \dots, x_{ip}), i= 1, \dots, n$ is a $(p+1)$ dimensional vector. In this paper, we focus on machine learning algorithms that act as a function on the input data. We seek a function $E\left[r\left(Y_{i}\right)\right] = f(\boldsymbol{x}_i) + \varepsilon_i,~i = 1, \dots, n,$ where $f : \mathbb{R}^p \to \mathbb{R}$, $Y$ is some response variable, $r$ is a known function and $\varepsilon$ is a random residual with zero mean and finite variance. A popular approach in actuarial science includes generalized linear models \cite{nelder1972generalized, ohlsson2010non}, where
\begin{equation}\label{eq:glm}
g(E[Y_i]) = \beta_0 + \sum_{j = 1}^{p} x_{ij}\beta_j = \boldsymbol{x}_i\boldsymbol{\beta},\quad i = 1, \dots, n,
\end{equation}
where $g$ is the link function, $\boldsymbol{\beta} = (\beta_0, \beta_1, \dots, \beta_p)'$, such that $f = g^{-1}\left(\boldsymbol{x}_i\boldsymbol{\beta}\right)$. This model is linear, ignoring interactions and non-linear transformations between features. 

Actuaries improve the pricing performance by adding hand-crafted features (through transformations or non-linear interactions). We will note the vector of these transformations with an asterisk, $\boldsymbol{x}^*$. A ratemaking model using the variables and the created features could have the same shape as \eqref{eq:glm}, but the vector of predictors would also include these transformations.

The novel approaches for ratemaking with machine learning involves flexible end-to-end models that transform the input data into predictions, see \cite{blier2021machine} for a review. In our experience of training end-to-end models P\&C insurance data, the optimal hyperparameters selected were not much more flexible than GLMs (not much depth in the neural network), meaning there were not enough labeled response variables (claim frequency or costs) to increase the model flexibility vastly. 

\begin{figure}[ht]
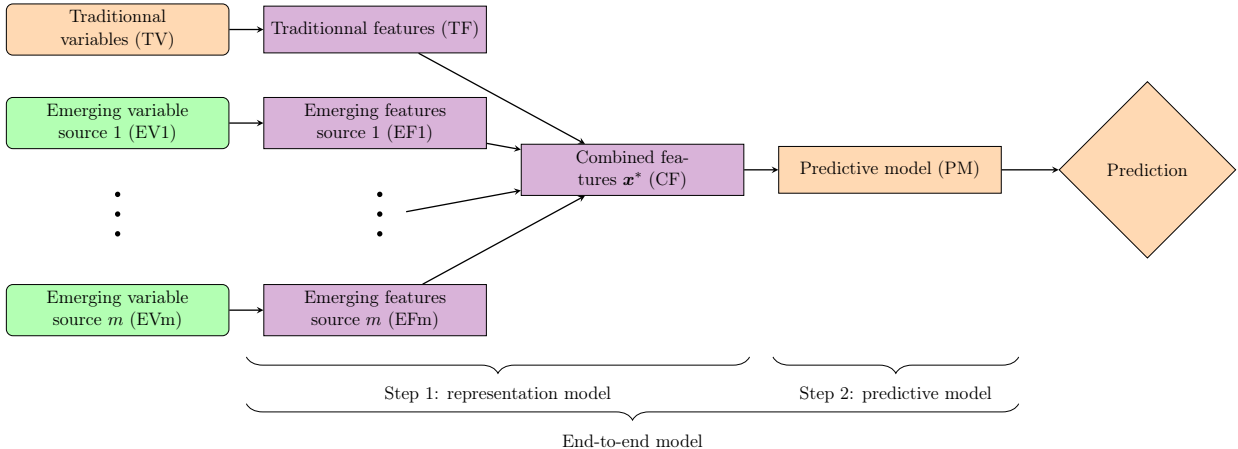

	\centering
	\include{img/process}
	\setlength{\belowcaptionskip}{-8pt} 
	\setlength{\abovecaptionskip}{-8pt} 
	\caption{Ratemaking process deconstructed. Orange: traditional ratemaking steps. Green: emerging data. Violet: representation steps.}\label{fig:process}
\end{figure}

In this paper, we detail a framework to automatically construct the feature vector $\boldsymbol{x}^*$ to use in a regression model. To accomplish this, we decompose the ratemaking process in two steps: 
\begin{itemize}
	\item[] step 1: a representation model to construct the embedding vector $\boldsymbol{x}^*$, and
	\item[] step 2: a regression model (typically a GLM) to model the claim frequency, claim severity or aggregate costs for an insurance contract, using $\boldsymbol{x}^*$ as inputs.
\end{itemize}
Figure \ref{fig:process} outlines the process of the approach. Our focus is solely on the representation model. Throughout the paper, we will provide different methods, from various emerging data, to construct a vectorial representation vector $\boldsymbol{x}^*$ that is useful. 

A useful representation transforms the input data such that the transformed feature improves the performance of a supervised learning model. If we create high-quality representations in step 1, it may be unnecessary to use more complex machine learning models in the regression model \cite{bengio2013representation}, only requiring GLMs. For instance, if a generalized additive model learns a polynomial relationship for a feature, then a good representation would also capture this polynomial relationship within $\boldsymbol{x}^*$, so a GLM using $\boldsymbol{x}^*$ would perform as well as a GAM using $\boldsymbol{x}$. With the same idea, a good representation could capture the non-linear transformations and interactions that tree-based models and deep neural networks do, all staying within the GLM framework. 

By using our approach, we can create more flexible models, in some instances, than a fully-supervised approach,\footnote{Although strategic pretraining of neural networks could replicate this increased flexibility for the fully-supervised approach.} by training the representation model using a larger dataset of unlabeled data (or labelled for another related task) all by remaining in a GLM framework. This is important since GLMs inherit maximum likelihood properties like unbiased and asymptotically normal parameter estimates, along with covariance from the inverse of the Fisher information matrix \cite{ohlsson2010non}. Neural networks typically do not reach global maximums, therefore not generating the maximum likelihood estimations. In these cases, we cannot perform hypothesis testing or build confidence intervals.

\subsection{Literature review}

Our unified approach to ratemaking with multisource data is influenced by \cite{collobert2008unified}, which provides a similar scope as this paper but for natural language processing (NLP). Our work fits in the field of representation learning, see \cite{bengio2013representation, goodfellow2016deep} for an overview on this topic. 

The representations in this paper use deep neural networks, a tool that many researchers in actuarial science have used in recent years. See \cite{blier2021machine} for a recent review of machine learning in P\&C ratemaking and reserving. In \cite{gao2018feature}, PCA (principal component analysis) and autoencoders are used to extract features from velocity-acceleration heatmaps created from telematics trips. These representations are then used in a regression model in \cite{gao2019claims}. Autoencoders and PCA are used in a tutorial to extract representations of categorical features in \cite{rentzmann2019unsupervised}. Then, \cite{blier2020encoding} compare PCA and autoencoders, then propose a convolutional neural network to extract spatial representations to classify the severity of car accidents. 

Our unified approach extracts features to use in a GLM model. The relationship between GLMs and deep neural networks was also recognized by \cite{wuthrich2019generalized}. Based on this relationship, they proposed the combined actuarial neural network (CANN). In this model, a GLM captures linear effects, and a neural network captures residual effects. We can interpret CANN as boosting the classical GLM model but leaving the GLM framework. \cite{henckaerts2020model} also use GLMs: the authors extract features of a gradient boosting machine using partial dependence effects and use them in a GLM. This approach aims to approximate a complex model with GLMs while retaining feature interpretability.

The idea of extracting intermediate representations is used in many fields; we cover a few below. \cite{bengio2003neural}, \cite{collobert2008unified} propose models to extract a distributed representation for words, a tool used in many applications of NLP and explained further in Section \ref{sec:nlp}. Representations of electronic health records are created in  \cite{miotto2016deep, rajkomar2018scalable}, generating simplified representations of patients used to predict medical outcomes. \cite{kadurin2017drugan} employ autoencoders to create representations used to identify molecular fingerprints with predefined anticancer properties. Even Youtube, a video streaming service, creates user profile representations with similar techniques to determine which video to recommend, see \cite{covington2016deep} for details.

\subsection{Outline of the paper}

The remainder of this paper is structured as follows. Section \ref{sec:data} presents the different formats and types of data that we will study in this paper. In Section \ref{sec:representation}, we structure modeling paradigms along a scale based on the degree of feature abstractness. We study the representation approach in detail in Section \ref{sec:construct} and explain different approaches to extract representations from non-vectorial data. Sections \ref{sec:nlp} to \ref{sec:geodata} provide more detail on our approach for emerging non-vectorial data in actuarial science, including text, images and spatial data respectively. Section \ref{sec:conclusion} concludes the paper. 


%% file: img/process.tex
\resizebox{\textwidth}{!}{
	\begin{tikzpicture}[node distance=2cm]
		\node (data) [data] {Traditionnal variables (TV)};
		\node (emerging1) [datae,below of=data] {Emerging variable source 1 (EV1)};
		\node (dots1) [below of = emerging1, scale = 3] {\rvdots};
		\node (emerging2) [datae,below of=dots1] {Emerging variable source $m$ (EVm)} ;
		
		\node (representation0) [process, right of = data, xshift=3.5cm] {Traditionnal features (TF)};
		\node (representation1) [process,right of = emerging1, xshift=3.5cm] {Emerging features source 1 (EF1)};
		\node (dots2) [right of = dots1, scale = 3, xshift=1.2cm] {\rvdots};
		\node (representation2) [process,right of = emerging2, xshift=3.5cm] {Emerging features source $m$ (EFm)};

		\node (representationc) [process,right of = representation1, xshift = 3.5cm, yshift=-1cm] {Combined features $\boldsymbol{x}^*$ (CF)};
		
		\node (predict) [processt, right of = representationc, xshift= 3.5cm] {Predictive model (PM)};
		\node (response) [decisiont, right of = predict, xshift = 3.5cm] {Prediction};
		
		\draw [arrow] (data) -- (representation0);
		\draw [arrow] (emerging1) -- (representation1);
		\draw [arrow] (emerging2) -- (representation2);
		\draw [arrow] (representation0) -- (representationc);
		\draw [arrow] (representation1) -- (representationc);
		\draw [arrow] (dots2) -- (representationc);
		\draw [arrow] (representation2) -- (representationc);
		\draw [arrow] (representationc) -- (predict);
		\draw [arrow] (predict) -- (response);
		
		\draw [decorate,decoration={brace,amplitude=10pt,mirror, raise = 0ex}] (2.75,-7) -- (13.5, -7) node[midway,yshift=-0.8cm]{Step 1: representation model};
		\draw [decorate,decoration={brace,amplitude=10pt,mirror, raise = 0ex}] (14,-7) -- (19.25, -7) node[midway,yshift=-0.8cm]{Step 2: predictive model};
		
		\draw [decorate,decoration={brace,amplitude=10pt,mirror, raise = 0ex}] (2.75,-8) -- (19.25, -8) node[midway,yshift=-0.8cm]{End-to-end model};

	\end{tikzpicture}
}

%% file: data.tex
\section{Emerging sources of data}\label{sec:data}

In this section, we present the emerging sources of data that have imminent potential in P\&C actuarial science, particularly for ratemaking. The data may be quantitative or qualitative. The data can also be structured into different formats, including vectorial data and image (matrix) data. Finally, the data can also be indexed to form a sequence. This section describes each and presents how they are stored for programming purposes.

Traditional data for P\&C insurance are described extensively in, e.g., \cite{parodi2014pricing}, \cite{denuit2020effectiveII}, \cite{ohlsson2010non}. We provide an overview of traditional data in Table \ref{tab:data}, along with emerging sources. 

\begin{table}[h]
	\begin{tabularx}{\textwidth}{lXX}
		\hline
		& Traditional                                                                    & Emerging                                                                                         \\ \hline
		All P\&C  & basic customer information (age, gender)                                       & granular customer information, IoT                                                               \\ \hline
		Auto      & type of car, miles driven                                                      & telematics, image of car, external data (weather, crime, traffic, vehicle identification number) \\ \hline
		Home      & construction characteristics, fire protection, distance to amenities, location & sensors, image of dwelling, external data (weather, crime, census)                               \\ \hline
		Liability & type of business, experience, credentials                                      & business reviews, lawsuit texts, underwriter comments                                            \\ \hline
	\end{tabularx}
	\caption{Traditional and emerging data for P\&C insurance}\label{tab:data}
\end{table}

\subsection{Types of data}\label{ss:qualitative}

Data can be quantitative or qualitative. Qualitative data consists of variables that can be measured on a discrete or continuous scale. We will often omit ordinal numbers and nominal numbers\footnote{Nominal numbers are used for identification and act as a one-to-one function between a number and a category or object, so typically represent categorical variables.} since mathematical operations do not make sense with these numbers. Since the data is numerical, it is straightforward to compute other numerical transformations such as summary statistics. Most statistical models exclusively deal with quantitative data since these numerical transformations are required to produce predictions. 


Many emerging variables are qualitative, also called categorical, which are usually classified into a fixed number of categories. We will use the term categorical data for the rest of this paper. Examples of categorical variables currently used in actuarial science include the following:
\begin{itemize}
	\item residential territory (zip / postal code, city / county, state / province);
	\item house siding material (wood, vinyl, stone, cement);
	\item car types (sedan, truck, sports).
\end{itemize}

Statistical models are not designed to take categorical data as features. Categorical variables must first be transformed into quantitative features. The common method to do this is called one-hot encoding in the machine learning nomenclature and is related to indicator features or dummy coding in statistical nomenclature.\footnote{The main difference between one-hot encoding and dummy coding is that one-hot encoding has no base category. Statistical models use dummy coding because using all categories yields non-unique solutions. This is not an issue for modern machine learning since the solutions are typically non-unique and yield local optima.} The one-hot encoding function transforms a single categorical variable into a vector that can be used in a statistical model. Consider a categorical variable that can take any category $j$ from 1 to $k$. The generated one-hot encoding is a vector $\boldsymbol{e}_j$ of length $k$ filled with zero, except at the $j$\textsuperscript{th} position where the value is 1. Consider a variable composed of five categories \{\texttt{Cat 1}, \texttt{Cat 2}, \texttt{Cat 3}, \texttt{Cat 4}, \texttt{Cat 5}\}, i.e. $k = 5$. An observation with \texttt{Cat 2} would generate a one-hot ending  $(0, 1, 0, 0, 0)'$, also noted $\boldsymbol{e}_2$. 

One-hot encoding is intuitive but has a few limitations when used in statistical models. First, as the number of categories increases, the dimensions of the encoding increases. This amplifies the curse of dimensionality. It also causes a loss in predictive power since no parameter is associated with new (unobserved) categories. Then, there is no measure of similarity between categories. Some categories may be ordinal, where differences between levels have hierarchical significance (a license to drive trucks implies a license to drive cars). In commercial insurance, the company classification may have levels of similarity. For instance, fast food companies and fine dining companies face similar risks and are more alike than construction or transportation companies. We will present methods to create more meaningful representations of categorical variables in Section \ref{sec:construct}. 

\subsection{Formats of data}

In this subsection, we present formats that individual variables can take for a single observation.

\subsubsection{Vectorial data}

The most straightforward data format is vectorial. This is the data type that spreadsheets handle efficiently. With vectorial data, the variables fit into a numeric vector. A consequence of having a dedicated element for each variable in the vector is that it does not naturally support non-fixed length variables. Since most emerging data sources have varying sizes (see later in this section for examples), these cannot be stored in tabular (record-oriented) datasets. 

Most machine learning models use vector data as inputs, including popular regression models like GLM/GAM, tree-based methods (CART, random forest, gradient boosting, see \cite{friedman2001elements}, \cite{denuit2020effectiveII}), and fully-connected neural networks. The unified approach for modeling non-fixed size variables in this paper is to introduce an intermediate step of vectorizing (creating vector features of fixed size) before being used for supervised regression. 

\subsubsection{Image (matrix) data}

\begin{figure}[ht]
	\centering
	\includegraphics[width=0.5\textwidth]{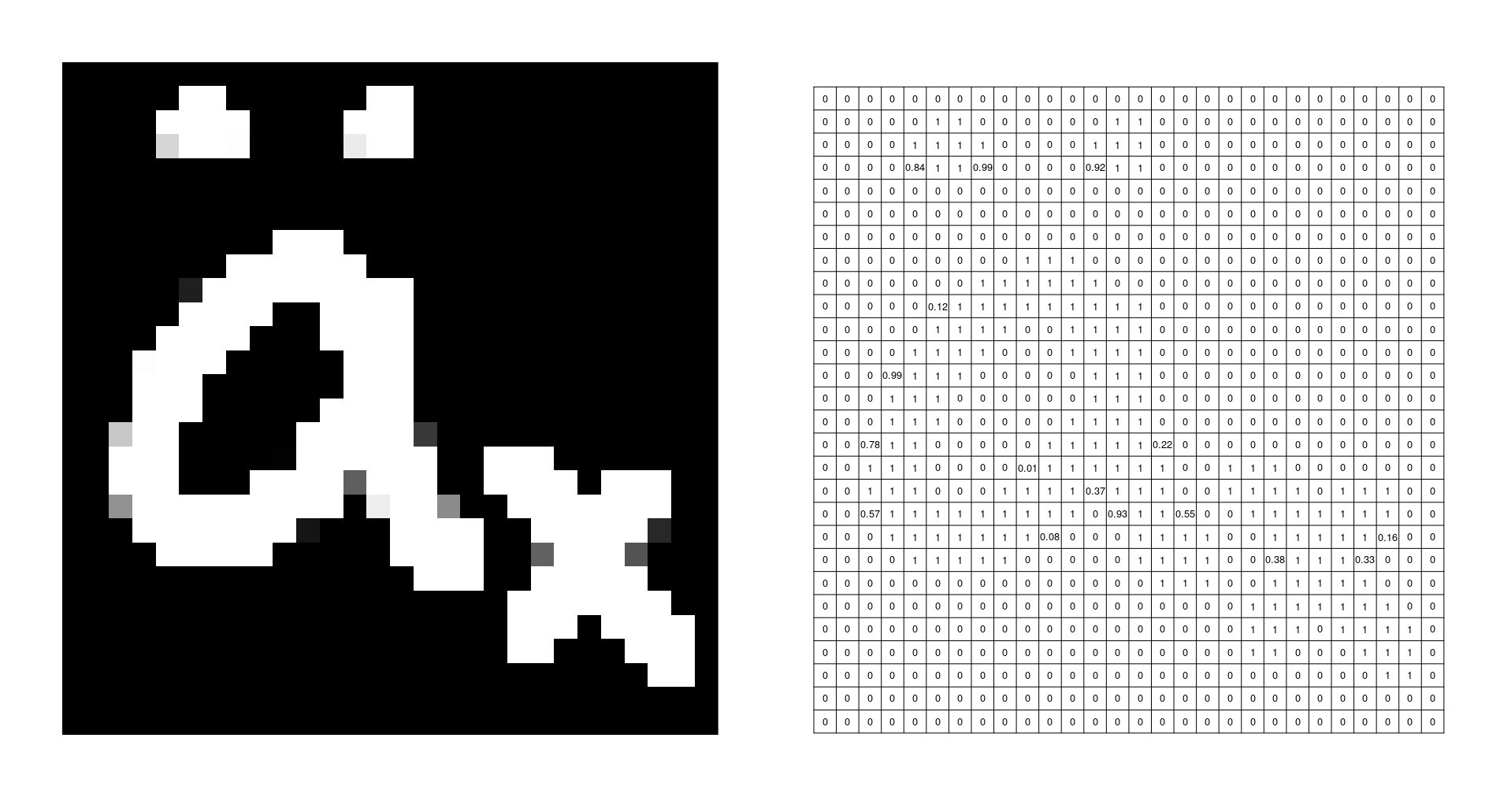}
	\setlength{\belowcaptionskip}{-8pt} 
	\setlength{\abovecaptionskip}{-8pt} 
	\caption{Left: grayscale image of a handwritten symbol, right: matrix representation of the channel}\label{fig:a-barre-x}
\end{figure}

This section briefly introduces the image format, although image handling will be revisited in Sections \ref{sec:construct} and \ref{sec:visual}. An image is a two-dimensional perception of a scene, usually containing one or more objects or subjects. In computers, they are projected and stored in a grid of pixels (assuming no compression). Consider an image with $N \times M$ pixels, for $\{N, M\} \in \mathbb{N}^2$. Each pixel will have associated values corresponding to an intensity (brightness). Images are stored in a computer as a matrix, where the intensity of each pixel is a integer number from 0 to 255 or floating point number from 0 to 1. A grayscale (black and white) image has one value (channel) associated with each pixel, with 1 corresponding to white and 0 corresponding to black. Figure \ref{fig:a-barre-x} presents an example of a 28$\times$28 grayscale image of a handwritten $\ddot{a}_x$ symbol along with the matrix of values. A typical color picture has three values (channels) for every pixel, representing intensities for the red, green and blue (RGB) channels. The corresponding tensor for a color picture will have size $M \times N \times 3$. 

\subsection{Indexing of data}

Indexing data leads to a sequence of variables, so the order of observations is important. We will use the notation $\{X_t, t \in T\}$ for realizations of the sequence and $|X|$ for the size of each observation. This includes time series, which are indexed at discrete times ($T \subseteq \mathbb{N}_0$) or continuous-time ($T \subseteq \mathbb{R}^+$). They can also be a sequence of ordered variables not indexed by time, which we will also note $T \subseteq \mathbb{N}_0$. Each variable could be of any format (qualitative or categorical) and of any type. Examples include 
\begin{itemize}
	\item individual reserving at the event level or snapshot (discrete-time interval) level;
	\item aggregate reserving (triangular data);
	\item stream data (including telematics), audio and video;
	\item textual data are sequences categorical data (words) from a large vocabulary;
	\item time series tracks the asset prices, and the sequence can be considered discrete or continuous.
\end{itemize}

We can use sequential data as inputs for many predictive tasks. On the one hand, a sequence can predict a single response variable (for instance, using a sequence of individual reserving payments to predict if a claim is open or closed). Models will have the shape 
$$E\left[r\left(Y_{i}\vert\boldsymbol{X}_{iT}\right)\right]  = f(\boldsymbol{X}_{iT}, \boldsymbol{x}_i) + \varepsilon_i, \quad i = 1, \dots, n,$$
where $\boldsymbol{X}_{iT}$ represents the sequence in the index set $T$, $f : \mathbb{R}^{m} \to \mathbb{R}$ is a predictive model, $m = |T| \cdot |X| + |\boldsymbol{x}_i|$, for $i = 1, \dots, n$. Models can also predict future observations in a sequence, like predicting the future payments in an individual reserving model in P\&C or generating the end of a sentence in NLP. These models use observations in $T$ to predict the unobserved observations in $\widehat{T}$:
$$ E\left[r\left(  \boldsymbol{X}_{i\widehat{T}}\vert\boldsymbol{X}_{iT}  \right)\right] = f(\boldsymbol{X}_{iT}, \boldsymbol{x}_i) + \varepsilon_{i}, \quad i = 1, \dots, n.$$

A numeric sequence of fixed length could be considered as vectorial data, with a column associated with each index in the sequence. Therefore, we can use any algorithm for vectorial data to model sequential data, ignoring the variables' sequential nature. However, statisticians typically use models that deal explicitly with sequential data, see Section \ref{sec:rep-seq}. 

%% file: representations.tex
\section{Data modeling paradigms}\label{sec:representation}

In this section, we classify different modeling paradigms based on the degree of feature abstractness, i.e. how abstract the features are, compared to the initial variables. The scale does not determine which method performs the best: this paper remains about data. Features with no abstraction use the variables themselves in an interpretable model, while features with the highest abstraction are implicit features obtained using black-box machine learning models. 

\subsection{Degree 0: variable-based learning}

The basic degree on the scale is to use the variables as features themselves. This means that quantitative variables are used as features directly, and qualitative variables are transformed into one-hot vectors and used directly. At this degree, the representation of the characteristics of an insured directly corresponds to its variables. Domain knowledge serves as an inspiration to identify new variables. From the process of Figure \ref{fig:process}, the data (TV, EV1, \dots, EVm) are the input to a linear predictive model (PM). 

\subsection{Degree 1: hand-crafted non-linear transformations \& interactions}

Insurers who want to improve predictive performance without the costly process of collecting new variables could create new features. At the first degree of abstraction, an insured's representation is its features and the new ones created. From the process of Figure \ref{fig:process}, we use data (TV, EV1, \dots, EVm) to construct features by hand (TF, EF1, \dots, EFm, CF) and use thse in a linear predictive model (PM). Consider again the GLM from \eqref{eq:glm}, where $g(E[Y_i])$ is a linear relationship between input features. The linear constraint implies that the model does not capture non-linear transformations of individual features or interactions between features.

For the model to consider non-linear transformations or interactions, actuaries can manually create features that capture them. This process is referred to as feature engineering. For instance, to capture the quadratic relationship of the variable \texttt{age}, the feature \texttt{age}\textsuperscript{2} must manually be added to the feature set. For the model to capture the effect of young male drivers, the actuary can create a feature \texttt{age} $\times$ $\boldsymbol{e}_j$ where $j$ is the category for male. 

We can also construct manual features from non-vectorial data. We could extract relevant statistics from raw telematics data, including average speed / acceleration, number of trips / hard breaks, and time of trips (peak or off-peak traffic). Credit scores are another example of such features, summarizing financial and credit data as a fiscal responsibility proxy.

This technique implies injecting \textit{a priori} knowledge within the data based on the domain knowledge of actuaries or underwriters. Much of actuarial work (and data science in general) is dedicated to creating these manual features. Although these features add much predictive performance on models, their construction are very time consuming. Finally, since there is an infinite candidate of non-linear transformation and interactions between features, it is improbable that the statistician will identify the best combinations by trial and error.

Three main problems arise from using degree 0 or degree 1 methods with non-vectorial data or vectorial data with lots of features (where $p$ is large). 
\begin{itemize}
	\item An individual feature may be insignificant when studying univariate statistics, but significant when studying multivariate statistics. 
	\item In variable-based learning, there is typically a feature selection step. The size $p$ of features from degree 0 or degree 1 may be very large, making feature selection tedious. While statistical techniques could select a subset of the variables to consider (see, e.g. \cite{pechon2018preliminary} to discard variable uncorrelated with the response variable or LASSO which implicitly selects significant features), variable selection methods have the following drawbacks:
	\begin{enumerate}
		\item Variable selection is computationally intensive, and greedy methods such as backward, forward and stepwise selection are non-optimal, especially for large $p$.
		\item Optimal variables may change if the modeling task changes, requiring repeated statistical analyses and computational variable selection. For example, if there are different models for two lines of business, variable selection must be repeated for each analysis. 
	\end{enumerate}
	\item The size $p$ of the vector of variables may not be constant for every observation, so \eqref{eq:glm} does not hold. For instance, an observation could have a telematics history, while another does not have one. Two insured will also drive a different amount, so the length of trips is different.
\end{itemize}

\subsection{Degree 2: representation learning}\label{sec:lev3} 

Representation learning is a method to circumvent the issues of degrees 0 and 1. The idea of degree 2 representations is to create dense and compact (shorter than the original dimension and non-zero entries) numerical vectors. These are called embeddings, and we construct them to capture structural and semantic information from variables. This degree of representations targets steps TF, EF1, \dots, EFm and CF of Figure \ref{fig:process}, building features automatically instead of manually. This degree is the focus of Section \ref{sec:construct}, and we provide specific examples in Sections \ref{sec:nlp} to \ref{sec:geodata}. 

\begin{figure}[ht]
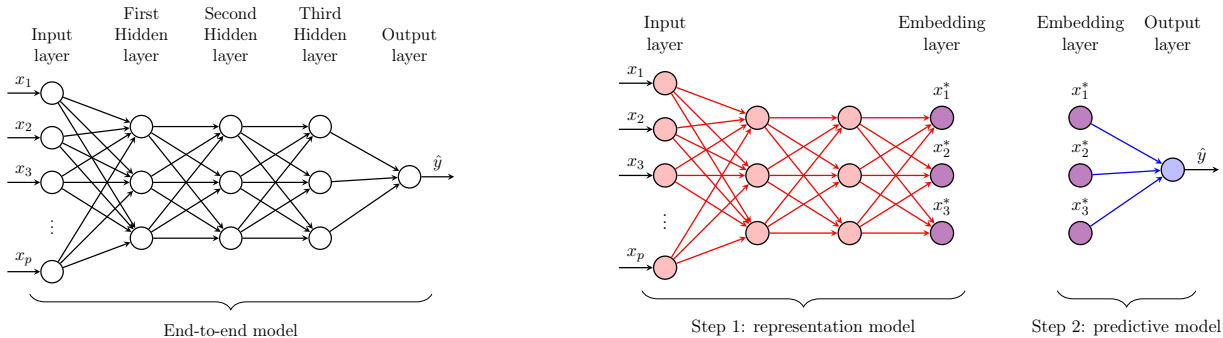

	\centering
	\begin{subfigure}[b]{0.37\textwidth}
		\centering
		\include{img/deep-nn-1}
		\setlength{\belowcaptionskip}{-8pt} 
		\setlength{\abovecaptionskip}{-8pt} 
		\caption{A deep neural network for regression}
		\label{fig:deep1}
	\end{subfigure}
	\hfill
	\begin{subfigure}[b]{0.5\textwidth}
		\centering
		\include{img/deep-nn-2}
		
		\setlength{\belowcaptionskip}{-8pt} 
		\setlength{\abovecaptionskip}{-8pt} 
		\caption{Deconstructed in two steps}
		\label{fig:deep2}
	\end{subfigure}
	\setlength{\belowcaptionskip}{-8pt} 
	\caption{A deep neural network for regression, interpreted as a feature generator and a GLM}\label{fig:deep}
\end{figure}

In modern machine learning (gradient boosting and neural networks), all variables are fed in a model that learns complex interactions between variables. This solves the first two points above (multivariate effects and variable selection). These models have excellent predictive performance but are often considered black boxes.\footnote{Techniques like LIME and SHAP exist to partially interpret the effects of inputs on the prediction.}
The models also have a large number of parameters, so diagnostic tools ignore model complexity. For this reason, we do not consider them as final pricing methods here and prefer using GLMs as predictive models. However, we can still use the intuitions behind deep neural network regressors and GLMs to explain embedding-based learning ideas. Consider a deep neural network for regression (see \cite{blier2021machine}, \cite{denuit2019effectiveIII} or \cite{wuthrich2017data} for details of deep neural network regression). The example in Figure \ref{fig:deep} has three hidden layers, and the last hidden layer contains three neurons. We can interpret the blue arrows and nodes as a GLM since $\hat{y} = g\left(\sum_{j = 1}^{3} w_{j}^{(3)}x_{j}^* + b^{(3)}\right)$, which has the same shape as \eqref{eq:glm}, and the features in the GLM are the hidden units of the final hidden layer of the neural network.\footnote{When the response variable is quantitative, we can interpret the neural network as being a regression GLM stacked on a feature map. When the response variable is categorical, we can interpret the model as being a logistic regression stacked on a feature map. Recall that although neural networks have the same shape as GLMs, they do not share the maximum likelihood properties that GLMs have.} This means that we can interpret the beginning of the network as a feature generator, applying non-linear transformations and interactions between input features to create a feature vector $\boldsymbol{x}^*$ that is ultimately used as inputs to a GLM.  


Using this intuition, we can keep the weights from the initial part of the model, creating embeddings of the input data, and train a new GLM for other tasks (see Figure \ref{fig:deep2}). We can construct these two models sequentially. The first step will be creating representations, also called pretraining embeddings, and the final step will be a GLM on the output of the first step. We explain how to construct embeddings in Section \ref{sec:construct}. Pretrained embeddings, constructed outside of the regression model, have many advantages for actuarial science, including:


\textbf{Transform non-vectorial data into vectorial data}. Much emerging data is not structured neatly; we cannot easily organize them in a row/column database. They are often unstructured formats, like text, images, sound, and geolocalised data. Statistical modeling of unstructured data is complex. An advantage of dense embeddings is that we can change the data type and format during construction. Many representation learning models are based on neural networks, with architectures that may handle multidimensional data (convolutional neural networks) and time-series data (recurrent neural networks). Using these models, we may extract structured representations of unstructured data and use this structured information in a traditional statistical model. 

\begin{figure}[ht]
		\centering
		\begin{subfigure}[b]{0.45\textwidth}
			\centering
			\begin{center}
				\begin{tabular}{ccc}
					{Category}            &       Dim. 1        &        Dim. 2         \\ \hline
					\texttt{Cat 1}          &         3.5         &          1.5          \\
					\texttt{Cat 2}          &         2.5         &           4           \\
					\texttt{Cat 3}          &          1          &          0.5          \\
					\texttt{Cat 4}          &          4          &           1           \\
					\texttt{Cat 5}          &          4          &          3.5          \\ \hline
					\textcolor{gray}{\texttt{Cat 6}} & \textcolor{gray}{3} & \textcolor{gray}{4.5}
				\end{tabular}
			\end{center}
			\caption{Embedding table}
			\label{fig:2dim-table}
		\end{subfigure}
		\hfill
		\begin{subfigure}[b]{0.45\textwidth}
			\centering
			\resizebox{0.5\textwidth}{!}{
				\begin{tikzpicture}
				\draw [thick, <->] (0,5) -- (0,0) -- (5,0);
				\node at (3.5, 1.5) {\LARGE \texttt{Cat 1}};
				\node at (2.5,4) {\LARGE \texttt{Cat 2}};
				\node at (1,0.5) {\LARGE \texttt{Cat 3}};
				\node at (4, 1) {\LARGE \texttt{Cat 4}};
				\node at (4,3.5) {\LARGE \texttt{Cat 5}};
				\node[gray] at (3,4.5) {\LARGE \texttt{Cat 6}};
				\end{tikzpicture}
			}
			\caption{Embedding plot}
			\label{fig:2dim-plot}
		\end{subfigure}
		\caption{Black: embeddings project one-hot encodings of dimension 5 into dense vectors of dimension 2. Gray: when adding a 6\textsuperscript{th} category, the embedding dimension remains two. }\label{fig:2dim}
	\end{figure}
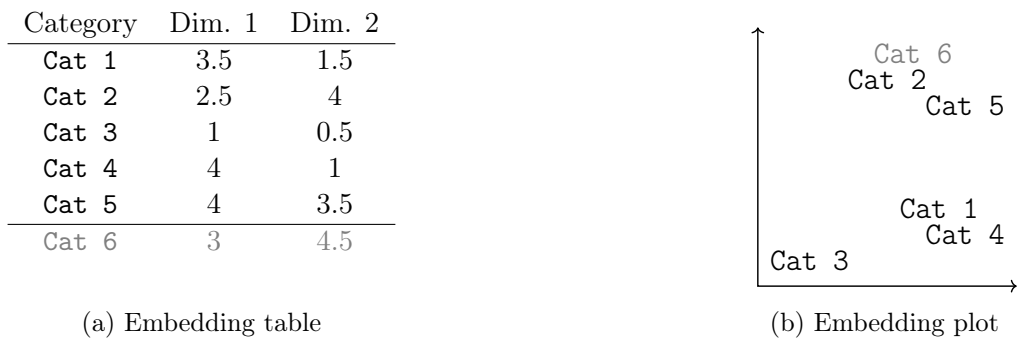

\textbf{Transform a categorical variable into a continuous one}. Representation models can compress categorical variables into dense vectors. When a categorical variable has many possible classes, the dimension of the one-hot encoding is also large (increasing $p$). The actuary chooses the embedding dimension $\ell$ with dense representations, typically selecting a dimension much smaller than the feature's cardinality. Fully-connected representations (see Section \ref{ss:fc}) are beneficial for categorical variables, which data scientists typically represent using one-hot encodings. Returning to our example of Section \ref{ss:qualitative}, consider a variables with $k = 5$ categories. The resulting one-hot vector has a dimension 5. A fully-connected neural network applied exclusively to this category would project this sparse vector into a dense vector with a dimension of our choice. In Figure \ref{fig:2dim}, we select a projection space in $\mathbb{R}^2$. A useful encoding will produce vectors of similar categories that lie in the same region of the vector space. For instance, \{2, 5\}, \{1, 4\} and \{3\} would each be clusters of similar categories. In addition, representations learned from datasets external to ratemaking could create vectors for new categories in the regression task (see \texttt{Cat 6} in gray from Figure \ref{fig:2dim}). With one-hot encodings, the new category would increase the dimension to 6, while embedding-based learning remains in 2 dimensions.

\textbf{Reduce data dimension.} Modern actuarial models use many features, but they may not all be statistically significant. Statisticians use feature selection to determine which features the model should use. Greedy algorithmic selection methods based on heuristics (e.g., backward, forward, stepwise) must train a model with many combinations of features and is not feasible for a large number of features. Expert selection is flawed because of expert bias, and features may accidentally be omitted, putting pressure on domain experts \cite{kuhn2019feature}. Representation learning methods allows to select the final dimension of the embeddings, typically decreasing $p$.

\textbf{Generates reusable representations.} Representations learned using techniques from Section \ref{sec:construct} are agnostic to the ultimate regression task. Once we create embeddings, we can use them in statistical models for related domains without modifying the embeddings. The actuary does not need to know how the embeddings are created.

\textbf{Learn non-linear transformations and interactions.} When using deep neural networks to construct embeddings, non-linear functions are used between each layer. This introduces non-linear transformations of combinations of input features. The representations are more abstract than the input features but capture useful transformations, and adding depth creates more representative embeddings \cite{bengio2013representation}. For models like GLMs (which do not create transformations and interactions), providing transformed features is essential. Furthermore, more flexible models like gradient boosting and neural networks may not be necessary for the regression task:  representation learning creates useful representations of the input data such that simpler models may be used in downstream tasks. 

\textbf{Decrease regression model complexity.} Related to the previous point, if we beak down the modeling process in two steps: a representation model followed by a regression model, then the complexity of the regression model does not depend on the complexity of the representation model. Consider a representation model to create embeddings $\boldsymbol{x}^* \in \mathbb{R}^{\ell}$ with a high number of parameters but which does use the response variable $Y$ during training. Then, a GLM regression model with $Y$ as a response variable and using $\boldsymbol{x^*}$ as input features will only have $\ell + 1$ parameters, irrespective of the representation model complexity.

In our experience of training predictive models, unsupervised embeddings produced statistically significant features even if hand-crafted features using the same underlying variables were already included in the model, meaning that unsupervised embeddings captured significant non-linear transformations and interactions. Representations add value even with data already used within models. 

\subsection{Degree 3: end-to-end learning}\label{sec:end-to-end}

The third degree of abstraction is implicit representations, meaning we never observe nor extract the representations. End-to-end learning removes the feature construction step, instead building a predictive model that will implicitly learn the appropriate representations. A conceptual difference is that the model architecture is specifically tailored to a task. It is more the model architecture that can be considered the representation, not the data itself. Examples of models at this degree include \cite{gabrielli2020neural}, who have specific network components to capture claim counts and claim severities, and \cite{delong2020collective}, who build a model with six neural networks to capture different sources of payments or recoveries within the claims process.

This degree of representation is the most flexible since it exits the linear predictive model framework. They can perform well in situations where domain knowledge leads to network architecture designs that capture unique characteristics of the predictive tasks (e.g. reserving in P\&C insurance) and in situations where a substantial volume of labeled data is available. When labeled data is limited, end-to-end models do not perform well in our experience. By exiting the GLM framework, quality-of fit statistics typically ignores model complexity and constructing confidence intervals that captures parameter uncertainty is tedious.



Table \ref{tab:summary-degree} presents a summary of properties for each degree. Since degree 2 representations separate the ratemaking process in two steps, we breakdown the properties for the representation model (construct) and the predictive model (use). Representation learning shares the simplicity of degree 0, while keeping (and sometimes improving) the flexibility of degree 3.

\begin{table}[ht]
	\centering
	\begin{tabular}{cccccc}
		                  Properties                    &     0      &     1      & 2 (construct) &  2 (use)   &     3      \\ \hline
		            Supports emerging data              & \texttimes & \texttimes &  \checkmark   &     N/A     & \checkmark \\
		      Supports non-linear transformations       & \texttimes & \checkmark &  \checkmark   & \checkmark & \checkmark \\
		     Requires \textbf{no} domain knowledge      & \checkmark & \texttimes &  \texttimes   & \checkmark & \texttimes \\
		Requires \textbf{no} machine learning knowledge & \checkmark & \checkmark &  \texttimes   & \checkmark & \texttimes \\
		              Not time consuming                & \checkmark & \texttimes &  \texttimes   & \checkmark & \texttimes \\
		          Stays in the GLM framework            & \checkmark & \checkmark &      N/A       & \checkmark & \texttimes
	\end{tabular}\caption{Summary of certain properties of models at different levels of abstraction}\label{tab:summary-degree}
\end{table}


%% file: img/deep-nn-1.tex
\resizebox{\textwidth}{!}{
	\begin{tikzpicture}[x=1cm, y=1cm, >=stealth]
	
	\foreach \m/\l [count=\y] in {1,2,3,missing,4}
	\node [every neuron/.try, neuron \m/.try, thick] (input-\m) at (0,2.5-\y) {};
	
	\foreach \m [count=\y] in {1,2,3}
	\node [every neuron/.try, neuron \m/.try, thick] (hidden1-\m) at (2,2-\y*1.25) {};
	
	\foreach \m [count=\y] in {1,2,3}
	\node [every neuron/.try, neuron \m/.try, thick] (hidden2-\m) at (4,2-\y*1.25) {};
	
	\foreach \m [count=\y] in {1,2,3}
	\node [every neuron/.try, neuron \m/.try, thick] (hidden3-\m) at (6,2-\y*1.25) {};
	
	\foreach \m [count=\y] in {1}
	\node [every neuron/.try, neuron \m/.try, thick] (output-\m) at (8,0.375-\y*0.75) {};
	
	\foreach \l [count=\i] in {1,2,3,p}
	\draw [<-, black, thick] (input-\i) -- ++(-1,0)
	node [above, midway] {$x_\l$};
	

	\foreach \l [count=\i] in {1}
	\draw [->, black, thick] (output-\i) -- ++(1,0)
	node [above, midway] {$\hat{y}$};
	
	\foreach \i in {1,...,4}
	\foreach \j in {1,...,3}
	\draw [->, black, thick] (input-\i) -- (hidden1-\j);
	
	\foreach \i in {1,...,3}
	\foreach \j in {1,...,3}
	\draw [->, black, thick] (hidden1-\i) -- (hidden2-\j);
	
	\foreach \i in {1,...,3}
	\foreach \j in {1,...,3}
	\draw [->, black, thick] (hidden2-\i) -- (hidden3-\j);
	
	\foreach \i in {1,...,3}
	\foreach \j in {1}
	\draw [->, black, thick] (hidden3-\i) -- (output-\j);
	
	\node [align=center, above] at (0,2) {Input \\ layer};
	\node [align=center, above] at (2,2) {First \\ Hidden \\ layer};
	\node [align=center, above] at (4,2) {Second \\ Hidden \\ layer};
	\node [align=center, above] at (6,2) {Third \\ Hidden \\ layer};
	\node [align=center, above] at (8,2) {Output \\ layer};
	
	\draw [decorate,decoration={brace,amplitude=10pt,mirror, raise = 0ex}] (-0.5,-3) -- (8.5, -3) node[midway,yshift=-0.8cm]{End-to-end model};
	
	\end{tikzpicture}
}

%% file: img/deep-nn-2.tex
\resizebox{\textwidth}{!}{
\begin{tikzpicture}[x=1cm, y=1cm, >=stealth]

\foreach \m/\l [count=\y] in {1,2,3,missing,4}
\node [every neuronr/.try, neuron \m/.try, thick] (input-\m) at (0,2.5-\y) {};

\foreach \m [count=\y] in {1,2,3}
\node [every neuronr/.try, neuron \m/.try, thick] (hidden1-\m) at (2,2-\y*1.25) {};

\foreach \m [count=\y] in {1,2,3}
\node [every neuronr/.try, neuron \m/.try, thick] (hidden2-\m) at (4,2-\y*1.25) {};

\foreach \m [count=\y] in {1,2,3}
\node [every neuronv/.try, neuron \m/.try, thick] (hidden3-\m) at (6,2-\y*1.25) {};

\foreach \m [count=\y] in {1,2,3}
\node [every neuronv/.try, neuron \m/.try, thick] (hidden4-\m) at (9,2-\y*1.25) {};
\foreach \m [count=\y] in {1}
\node [every neuronb/.try, neuron \m/.try, thick] (output-\m) at (11,0.375-\y*0.75) {};

\foreach \l [count=\i] in {1,2,3,p}
\draw [<-, thick] (input-\i) -- ++(-1,0)
node [above, midway] {$x_\l$};

%
%
\foreach \l [count=\i] in {1,2,3}
\node [above] at (hidden3-\i.north) {$x^*_{\l}$};

\foreach \l [count=\i] in {1,2,3}
\node [above] at (hidden4-\i.north) {$x^*_{\l}$};

\foreach \l [count=\i] in {1}
\draw [->, thick] (output-\i) -- ++(1,0)
node [above, midway] {$\hat{y}$};

\foreach \i in {1,...,4}
\foreach \j in {1,...,3}
\draw [->, red, thick] (input-\i) -- (hidden1-\j);

\foreach \i in {1,...,3}
\foreach \j in {1,...,3}
\draw [->, red, thick] (hidden1-\i) -- (hidden2-\j);

\foreach \i in {1,...,3}
\foreach \j in {1,...,3}
\draw [->, red, thick] (hidden2-\i) -- (hidden3-\j);

\foreach \i in {1,...,3}
\foreach \j in {1}
\draw [->, blue, thick] (hidden4-\i) -- (output-\j);

\node [align=center, above] at (0,2) {Input \\ layer};
\node [align=center, above] at (6,2) {Embedding \\ layer};

\node [align=center, above] at (9,2) {Embedding \\ layer};
\node [align=center, above] at (11,2) {Output \\ layer};

\draw [decorate,decoration={brace,amplitude=10pt,mirror, raise = 0ex}] (-0.5,-3) -- (6.5, -3) node[midway,yshift=-0.8cm]{Step 1: representation model};
\draw [decorate,decoration={brace,amplitude=10pt,mirror, raise = 0ex}] (8.5,-3) -- (11.5, -3) node[midway,yshift=-0.8cm]{Step 2: predictive model};

\end{tikzpicture}
}

%% file: constructing.tex
\section{Learning representations}\label{sec:construct}

An embedding is a dense representation of \textit{input features} to perform a \textit{predictive task}. This section details the four steps to construct a representation model to generate embeddings (colored violet). There are two components in a representation model: the encoder (colored red) and the decoder or predictor (colored blue). The first step is to construct the encoder, an architecture that projects raw variables into embeddings. The second step is to construct the decoder, which will determine which domain knowledge the embeddings capture. The encoder and decoder will perform the processes (TF, EF1, \dots, EFm) in the process from Figure \ref{fig:process}. The third step combines representations from multiple data sources (step CF of Figure \ref{fig:process}), while the fourth step will be to evaluate the representations. 

The training procedures for neural networks (estimating the model's parameters) are out of this paper's scope since any backpropagation algorithm can perform this task. The unfamiliar reader can refer to \cite{goodfellow2016deep} for details on training neural networks.

\subsection{Step 1: the encoder}

The first step in constructing representations is to set up the compression architecture by answering \textit{what goes in the model} and \textit{what are the hidden steps in the encoder}. For this step, neural networks are a popular choice since they offer a lot of architectural flexibility and predictive performance. We explain three methods to extract vectorial representations for emerging data. For each method, we present the particular operations that capture the unique data characteristics, a topological framework, and explain how the variables become embeddings.

\subsubsection{Representations of vectorial data with fully-connected neural networks}\label{ss:fc}


The basic deep neural networks are fully-connected. This is the model presented in Figure \ref{fig:deep2}, and supports vectorial data as input. Following \cite{blier2021machine}, let $h^{(L-1)}_j, j = 1, \dots, J^{(L-1)}$ be the last hidden layer, with $L$ being the number of layers and $J^{(L-1)}$ is the size of the last hidden layer. Then, we can select $\boldsymbol{x}^* = \boldsymbol{h}^{(L-1)}$ as features generated by the function $f: \mathbb{R}^{p} \to \mathbb{R}^{J^{(L-1)}}$ which transforms the input features $\boldsymbol{x}$ into the embedding vector $\boldsymbol{x}^*$.


\subsubsection{Representations of sequences with recurrent neural networks}\label{sec:rep-seq}

\begin{figure}[ht]
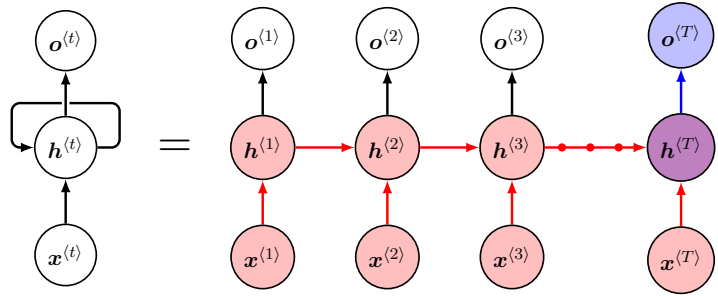

	\begin{center}
		\include{img/rnn}
	\end{center}
	\setlength{\belowcaptionskip}{-8pt} 
	\setlength{\abovecaptionskip}{-16pt} 
	\caption{A simple RNN architecture. Left: recurrent representation of the architecture. Right: unrolled representation.}
	\label{fig:rnn}
\end{figure}

Recurrent neural networks naturally capture the sequential nature of data. The network iteratively updates its state with each element of a sequence, see Figure \ref{fig:rnn} for a simple example. Let the superscript $\langle t \rangle$ denotes state at time $t, t = 1, \dots, T$. Let $\rnnvect{x}{t}\in \mathbb{R}^{p}$ be the vector of input features at time $t$. Let $\rnnvect{h}{t}\in \mathbb{R}^{\ell}$ be the vector of the (hidden) cell state at time $t$ and $\rnnvect{o}{t}\in \mathbb{R}^{J^{(o)}}$ the output features at time $t$, for $t = 1, \dots, T$. 

The basic idea behind RNNs is that the hidden state $\rnnvect{h}{t}$ depends on the previous hidden state $\rnnvect{h}{t-1}$ and the current input vector $\rnnvect{x}{t}$. This enables some sort of memory, so the hidden states $\rnnvect{h}{t}$ depend on all previous observations in the sequence $\rnnvect{x}{u}, u = 1, \dots, t$. The relationship between the previous hidden state and the new input vector is $\rnnvect{h}{t} = g^{h}\left(\rnnvect{z}{t} \right)$, where $g^h$ is an activation function and
$$\rnnvect{z}{t} = \underbrace{\rnnvect{x}{t}\boldsymbol{W}^{[x]} + \parvect{b}{x}}_{\text{new information}} + \underbrace{\rnnvect{h}{t-1}\boldsymbol{W}^{[h]}+ \parvect{b}{h}}_{\text{memory}},$$
where the first term depends on $\rnnvect{x}{t}$ and represents new information, while the second term depends on $\rnnvect{h}{t-1}$ and represents the memory from past states, and $\boldsymbol{W}^{[x]} \in \mathbb{R}^{p\times \ell}, \parvect{b}{x} \in \mathbb{R}^{\ell}, \boldsymbol{W}^{[h]} \in \mathbb{R}^{\ell\times \ell}, \parvect{b}{h} \in \mathbb{R}^{\ell}$ are trainable weight parameters. Finally, one or more fully connected layers are usually used to produce an output prediction. For one layer, the output score is $\rnnvect{z}{t, o} = \rnnvect{h}{t}\boldsymbol{W}^{[o]} + \parvect{b}{o}$, where $\boldsymbol{W}^{[o]} \in \mathbb{R}^{\ell \times J^{(o)}}, \parvect{b}{o}\in \mathbb{R}^{J^{(0)}}$, and the output value is $\rnnvect{o}{t} = g^{o}\left(\rnnvect{z}{t, o}\right)$, where $g^{o}$ is the output activation function.

The hidden state vectors $\rnnvect{h}{t}, t = 1, \dots, T$ can be considered as representations of the sequence because a single hidden state represents all input features from prior observations in the sequence. 
The architecture from Figure \ref{fig:rnn} has $T$ hidden states, then we can select the embedding representation as the hidden state after the most recent observation in the sequence, i.e. $\boldsymbol{x}^* = \rnnvect{h}{T}$. 
Since the dimension of the hidden layer $\rnnvect{h}{t}, t = 1, \dots, T$ is constant (chosen as a hyperparameter of the neural network), the embedding is always of fixed length. Therefore, for different lengths $T$, the representation can be considered as vectorial data, i.e. the representation model is $f: \mathbb{R}^{T\times p} \to \mathbb{R}^{\ell}.$

Our simple recurrent neural network's hidden state was the sum of a hidden state and new information, but modern RNN models use more complex hidden states. For instance, long short term memory (LSTM, \cite{hochreiter1997long}) and gated recurrent units (GRU, \cite{cho2014learning}) include gates to determine to what extent new information should be included or excluded in future hidden states. 

%

\subsubsection{Representations of images with convolutional neural networks}

\begin{figure}[ht]
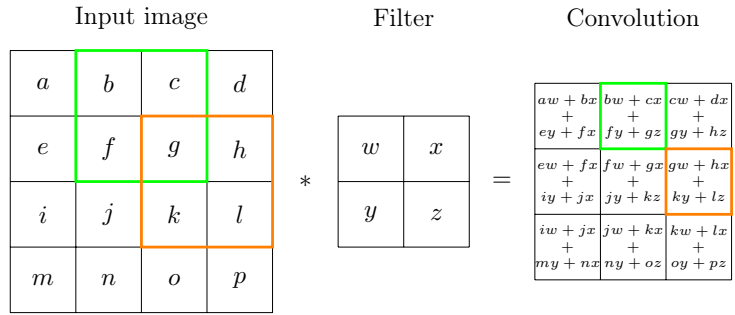

	\centering
	\include{conv-op}
	\setlength{\belowcaptionskip}{-5pt} 
	\setlength{\abovecaptionskip}{-8pt} 
	\caption{Example convolution for image size $4\times 4$ and filter size $2\times 2$}\label{fig:conv-op}
	
\end{figure}

Convolutional neural networks rely on the convolution operation to capture order and distance from text, sound and image data. The convolution operation of a filter applied to these data sources creates local features. These features are useful since they capture inherent meaning from sections of the image or of the sequence. Figure \ref{fig:conv-op} provides an illustration of this operation. For details on the discrete convolution for images, see \cite{goodfellow2016deep} or \cite{dumoulin2016guide}.

\begin{figure}[ht]
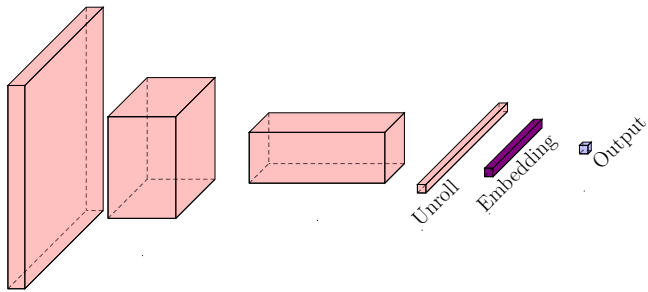

	\centering
	\include{conv-extract}
	\setlength{\belowcaptionskip}{-5pt} 
	\caption{Creating an embedding with convolutional neural networks}\label{fig:conv}
\end{figure}

Consider a color image of dimension $N \times M$. To create vectorial representations of image data with CNNs, we first pass the input image through convolution layers. Then, we transform a three dimensional feature map into one dimension by taking the  convolutional representation and unrolling the feature map into a vector. That is, after the convolution operations, the feature space is a square cuboid with three dimensions $(q, q, c)$, where $q, c \in \mathbb{N}$. This feature space is \textit{unrolled} (from left to right, top to bottom, front to back) into a vector $\boldsymbol{h} \in \mathbb{R}^{q \cdot q \cdot c}$. This unrolled vector can be followed by fully-connected layers and the last fully-connected layer before the prediction is considered the embedding vector $\boldsymbol{x}^*$. The purpose of adding a fully-connected layer after the unroll stage is to reduce the spatial autocorrelation that is still present in the last convolution feature map. Figure \ref{fig:conv} presents an example with a single fully-connected layer between the unrolled representation and the embedding. Then, $\boldsymbol{x}^* = g(\boldsymbol{z})$, where $g$ is the activation function
$\boldsymbol{z} = \boldsymbol{hW} + \boldsymbol{b}$, with $\boldsymbol{W} \in \mathbb{R}^{(q \cdot q \cdot c) \times \ell}$ and $\boldsymbol{b} \in \mathbb{R}^{\ell}$.

Figure \ref{fig:conv} presents a CNN architecture to create vectorial embeddings from images. We can extract a function $f: \mathbb{R}^{N \times M \times 3} \to \mathbb{R}^{\ell}$ that maps a color image to an embedding vector $\boldsymbol{x}^*$ of dimension $\ell$.

We note that we can also represent sequences with one-dimensional convolutional neural networks. These models are used for text modeling and telematics modeling \cite{gao2019convolutional} but RNNs naturally capture sequential properties.

\subsection{Step 2: the decoder}

The decoding or predictive step is vital since it determines the knowledge the representation model induces in the embeddings. This subsection presents a few approaches.



\subsubsection{Representations of itself: autoencoders and unsupervised learning}

An unsupervised method to construct representations is to build a model to predict the input features themselves. Using a large quantity of information, we first construct vectors $\boldsymbol{x}^*$ projecting input data in a latent space of smaller dimension. Undercomplete (bottleneck) autoencoders accomplish dimension reduction by building an encoder and a decoder. The encoder compresses input features into a smaller dimension than the input dimension, and the decoder attempts to project the compressed data back by reconstructing the input features. Let $f: \mathbb{R}^p \rightarrow \mathbb{R}^\ell$ be the encoder and $g: \mathbb{R}^\ell \rightarrow \mathbb{R}^p,$ be the decoder, where $p$ is the input dimension and $\ell$ is the latent space (embedding, or representation) dimension. To perform dimension reduction, we choose $l\ll p$. To build this model, we seek an encoder $f$ and a decoder $g$ such that the reconstruction error is minimized, 
\begin{equation}\label{eq:reconstruction}
\argmin_{g, f} \sum_{i = 1}^{n}\sum_{j = 1}^{p}(g(f(x_{i, j})) - x_{i, j})^2.
\end{equation}
Multiple classes of functions exist for $f$ and $g$ to perform this optimization. 

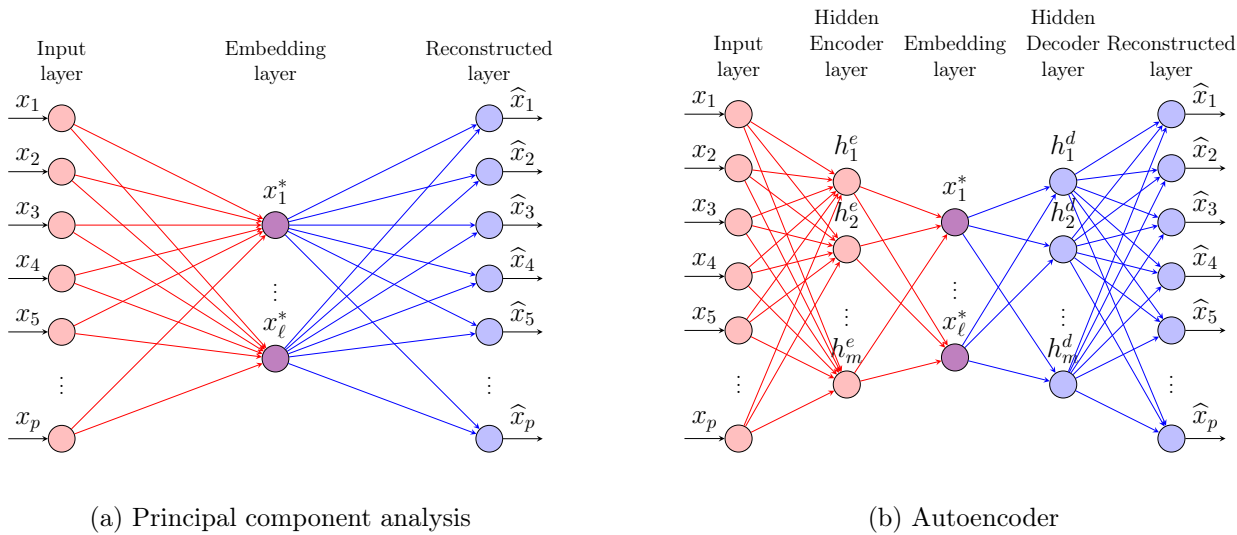
\begin{figure}[ht]
	\centering
	\begin{subfigure}[b]{0.45\textwidth}
		\centering
		\input{img/pca.tex}
		\caption{Principal component analysis}
		\label{fig:ae_shape}
	\end{subfigure}
	\hfill
	\begin{subfigure}[b]{0.45\textwidth}
		\centering
		\input{img/autoencoder.tex}
		\caption{Autoencoder}
		\label{fig:ae_neurons}
	\end{subfigure}
	\caption{Bottleneck architectures of PCA and autoencoders}
	\label{fig:one-dim}
\end{figure}

In principle component analysis, $\boldsymbol{x}^*_i = f\left(\boldsymbol{x}_i\right) = W\boldsymbol{x}_i, i= 1, \dots, n$, where $W \in \mathbb{R}^{p\times \ell}$ is an orthonormal matrix corresponding to the first eigenvectors of the covariance matrix of $\boldsymbol{x}_i$. Although proofs for principal components usually rely on a variance maximization objective, we can also show that the principal components minimize the reconstruction error in \eqref{eq:reconstruction} subject to $W$ being an orthonormal basis. The embedding layer are the principal components of the data; see Figure \ref{fig:ae_shape}. 

The representation ability of PCA is restricted to linear combinations. With deep learning, the representations are more flexible. Adding more depth may create more abstract representations --typically more useful in models-- but are harder to train \cite{bengio2013representation}.

Autoencoders optimize (\ref{eq:reconstruction}) using a fully-connected neural network as an encoder and a fully-connected neural network as a decoder. When $\ell \ll p$, encoders \textit{compress} input data to a lower dimension layer, and the decoder attempts to reconstruct (decompress) the input data. If the compressed information is mostly reconstructed, the compressed features retain the input data's useful information. Figure \ref{fig:ae_neurons} contains a graphical representation of an autoencoder with $p$ input features, one hidden layer of $m$ neurons and an embedding layer of size $\ell$. The values $h^e_j$ and $h^d_j, j = 1, \dots, m$ are hidden layers of the encoder and of the decoder.


\subsubsection{Representations of other tasks: transfer learning}\label{ss:transfer}

Transfer learning creates representations of similar tasks. If the tasks are similar, the representations will be useful in both cases. Consider a car insurer that launches a new business line in motorcycle insurance. The insurance company cannot create motorcycle drivers' representations due to the lack of data, but they can create car drivers' representations. Since car insurance claim histories include domain knowledge on the risk level, the car driver representation will be useful as a motorcycle driver representation. Transfer learning induces domain information from related tasks into a representation \cite{caruana1997multitask}.

Embeddings extracted from transfer learning are typically the last hidden layer before the output layer, see Figure \ref{fig:deep}. This layer is the farthest from the input layer (thus highly non-linear), and close to the prediction such that the hidden layer contains useful information for prediction.

\subsection{Step 3 (optional): combining representations}

Now that we have constructed embeddings of many emerging sources of data, we must combine these embeddings into one feature vector that will be used in a predictive model. To combine representations, we can use techniques from encoder and decoder techniques above.

For example, consider a model with a vectorial representation of dimension $\ell_1$, a sequential representation of dimension $\ell_2$ and an image representation of dimension $\ell_3$. Concatenating these representations generates a single vector of dimension $\ell_1 + \ell_2 + \ell_3$. We can optionally mix these embeddings by adding fully-connected layers after the concatenation and train a new combined embedding with transfer learning (\ref{ss:transfer}) or by constructing an autoencoder with the concatenated representation. We could also combine the three pre-trained encoder models into a single encoder and fine-tune the weights for a combined decoder.

\subsection{Step 4 (optional): evaluating representations}

The last optional step is to evaluate the representations. There are two types of methods to perform this task. The first is intrinsic, where we determine if the representations make sense. This method usually consists of comparing the distance between two representations. A popular distance measure is the cosine distance between vectors,
\begin{equation}\label{eq:cosine}
\text{cosine}(\boldsymbol{u}, \boldsymbol{v}) = \frac{\boldsymbol{u}\cdot \boldsymbol{v}}{||\boldsymbol{u}|| ~ ||\boldsymbol{v}||},
\end{equation}
where $||\boldsymbol{u}||$ is the Euclidean norm. We can perform intrinsic evaluations of representations by selecting a vector, computing the cosine distance with other vectors and studying the observations with the smallest and largest cosine distances. The representations make sense if small cosine distances occur with similar observations, and large cosine distances occur with dissimilar observations. Researchers use benchmark datasets to evaluate the quality of word embeddings in fields like NLP, comparing cosine distances with human-assigned similarity scores like wordsim353 \cite{finkelstein2002placing}. 

The second type of representation evaluation is extrinsic evaluations. In this method, we use the representations in downstream tasks, i.e. use them as features in a predictive model. When comparing two representation models, the one who performs the best on the ratemaking prediction is preferred. A data scientist that constructs representations should select downstream tasks beforehand and select a performance threshold at which the embeddings are acceptable for use.

%% file: img/rnn.tex
\resizebox{!}{0.17\textheight}{
	\begin{tikzpicture}[item/.style={circle,draw,thick,align=center},itemc/.style={item,on chain,join}]

		\node[item, fill=red!25] (A1) {$\rnnvect{h}{1}$};
		\node[item, right of = A1, xshift=3em, fill=red!25] (A2) {$\rnnvect{h}{2}$};
		\node[item, right of = A2, xshift=3em, fill=red!25] (A3) {$\rnnvect{h}{3}$};
		\node[item, right of = A3, xshift=5em, fill = violet!50] (AT) {$\rnnvect{h}{T}$};
		\draw[very thick,-latex, red] (A1) -- (A2);
		\draw[very thick,-latex, red] (A2) -- (A3);
		\draw[very thick,-latex, red] (A3) -- (AT);

		\node[left=1em of A1,scale=2] (eq) {$=$};
		\node[left=2em of eq,item] (AL) {$\rnnvect{h}{t}$};
		\path (AL.west) ++ (-1em,2em) coordinate (aux);
		\draw[very thick,-latex,rounded corners] (AL.east) -| ++ (1em,2em) -- (aux) 
		|- (AL.west);
		\foreach \X in {1,2,3} 
		{\draw[very thick,-latex] (A\X.north) -- ++ (0,2em)	node[above,item] (h\X) {$\rnnvect{o}{\X}$};
			\draw[very thick,latex-, red] (A\X.south) -- ++ (0,-2em) node[below,item, black, fill=red!25] (x\X) {$\rnnvect{x}{\X}$};}
		
		\draw[very thick,-latex, blue] (AT.north) -- ++ (0,2em) node[above,item, black, fill=blue!25] (hT) {$\rnnvect{o}{T}$};
		\draw[very thick,latex-, red] (AT.south) -- ++ (0,-2em) node[below,item, black, fill=red!25] (xT) {$\rnnvect{x}{T}$};
		
		\draw[white,line width=0.8ex] (AL.north) -- ++ (0,1.9em);
		\draw[very thick,-latex] (AL.north) -- ++ (0,2em) node[above,item] {$\rnnvect{o}{t}$};
		\draw[very thick,latex-] (AL.south) -- ++ (0,-2em) node[below,item] {$\rnnvect{x}{t}$};
		\path (A3) -- (AT) node[midway,scale=2.5,font=\bfseries, red] {\dots}; https://tex.stackexchange.com/questions/494139/how-do-i-draw-a-simple-recurrent-neural-network-with-goodfellows-style
	\end{tikzpicture}
}

%% file: conv-op.tex
\resizebox{0.6\textwidth}{!}{
\begin{tikzpicture}
	\draw[step=1.0,black,thin] (0,0) grid (4,4);
	\foreach \m [count=\y] in {a, b, c, d}
	\node at (\y - 0.5, 3.5) {$\m$};
	\foreach \m [count=\y] in {e, f, g, h}
	\node at (\y - 0.5, 2.5) {$\m$};
	\foreach \m [count=\y] in {i, j, k, l}
	\node at (\y - 0.5, 1.5) {$\m$};
	\foreach \m [count=\y] in {m, n, o, p}
	\node at (\y - 0.5, 0.5) {$\m$};
	
	\node at (4.5, 2) {$\ast$};
	
	\draw[step=1.0,black,thin] (5,1) grid (7,3);
	\foreach \m [count=\y] in {w, x}
	\node at (\y + 4.5, 2.5) {$\m$};
	\foreach \m [count=\y] in {y, z}
	\node at (\y + 4.5, 1.5) {$\m$};
	
	\node at (7.5, 2) {$=$};
	
	\draw[step=1.0,black,thin,yshift=0.5cm] (8,0) grid (11,3);
	
	\node at (8.5, 3.25) {\tiny $aw +bx$};
	\node at (8.5, 3) {\tiny $+$};
	\node at (8.5, 2.75) {\tiny $ey+fx$};
	
	\node at (9.5, 3.25) {\tiny $bw+cx$};
	\node at (9.5, 3) {\tiny $+$};
	\node at (9.5, 2.75) {\tiny $fy+gz$};
	
	\node at (10.5, 3.25) {\tiny $cw+dx$};
	\node at (10.5, 3) {\tiny $+$};
	\node at (10.5, 2.75) {\tiny $gy+hz$};
	
	\node at (8.5, 2.25) {\tiny $ew +fx$};
	\node at (8.5, 2) {\tiny $+$};
	\node at (8.5, 1.75) {\tiny $iy+jx$};
	
	\node at (9.5, 2.25) {\tiny $fw+gx$};
	\node at (9.5, 2) {\tiny $+$};
	\node at (9.5, 1.75) {\tiny $jy+kz$};
	
	\node at (10.5, 2.25) {\tiny $gw+hx$};
	\node at (10.5, 2) {\tiny $+$};
	\node at (10.5, 1.75) {\tiny $ky+lz$};
	
	\node at (8.5, 1.25) {\tiny $iw +jx$};
	\node at (8.5, 1) {\tiny $+$};
	\node at (8.5, 0.75) {\tiny $my+nx$};
	
	\node at (9.5, 1.25) {\tiny $jw+kx$};
	\node at (9.5, 1) {\tiny $+$};
	\node at (9.5, 0.75) {\tiny $ny+oz$};
	
	\node at (10.5, 1.25) {\tiny $kw+lx$};
	\node at (10.5, 1) {\tiny $+$};
	\node at (10.5, 0.75) {\tiny $oy+pz$};
	
	\draw [step=2.0,green, very thick] (1,2) rectangle (3,4);
	\draw [step=2.0,green, very thick] (9,2.5) rectangle (10,3.5);
	
	\draw [step=2.0,orange, very thick] (2,1) rectangle (4,3);
	\draw [step=2.0,orange, very thick] (10,1.5) rectangle (11,2.5);
	
	\node at (2, 4.5) {Input image};
	\node at (6, 4.5) {Filter};
	\node at (9.5, 4.5) {Convolution};
	
\end{tikzpicture}
}

%% file: conv-extract.tex
\resizebox{0.35\textwidth}{!}{
	\begin{tikzpicture}
	\tikzstyle{connection}=[ultra thick,every node/.style={sloped,allow upside down},draw=\edgecolor,opacity=0.7]
	\tikzstyle{copyconnection}=[ultra thick,every node/.style={sloped,allow upside down},draw={rgb:blue,4;red,1;green,1;black,3},opacity=0.7]
	
	\path[use as bounding box] (0, -2.5) rectangle (10, 4);

	\pic[shift={(0,0,0)}] at (0,0,0) 
	{Box={
			name=conv1,
			caption= ,
			xlabel={{, }},
			zlabel=,
			fill=red,
			opacity=0.25,
			height=24,
			width=2,
			depth=24
		}
	};
	
	\pic[shift={(1.5,0,0)}] at (conv1-east) 
	{Box={
			name=conv2,
			caption= ,
			xlabel={{, }},
			zlabel=,
			fill=red,
			opacity=0.25,
			height=12,
			width=8,
			depth=12
		}
	};
	
	\pic[shift={(1.5,0,0)}] at (conv2-east) 
	{Box={
			name=conv3,
			caption= ,
			xlabel={{, }},
			zlabel=,
			fill=red,
			opacity=0.25,
			height=6,
			width=16,
			depth=6
		}
	};

	\pic[shift={(1.5,0,0)}] at (conv3-east) 
	{Box={
			name=fc1,
			caption= ,
			xlabel={{, }},
			zlabel=\Large Unroll,
			fill=red,
			opacity=0.25,
			height=1,
			width=1,
			depth=25
		}
	};
	
	\pic[shift={(1,0,0)}] at (fc1-east) 
	{Box={
			name=soft1,
			caption=,
			xlabel={{" ","dummy"}},
			zlabel=\Large Embedding,
			fill=violet,
			opacity=1,
			height=1,
			width=1,
			depth=15
		}
	};

	\pic[shift={(1.5,0,0)}] at (soft1-east) 
	{Box={
			name=soft1,
			caption=,
			xlabel={{" ","dummy"}},
			zlabel=\Large Output,
			fill=blue,
			opacity=0.25,
			height=1,
			width=1,
			depth=1
		}
	};
	
	\end{tikzpicture}
}

%% file: img/pca.tex
\resizebox{\textwidth}{!}{
	\begin{tikzpicture}[x=1cm, y=1cm, >=stealth]
	
	
	\foreach \m/\l [count=\y] in {1,2,3,4, 5, missing,6}
	\node [every neuronr/.try, neuron \m/.try] (input-\m) at (0,2.5-\y) {};
	
	\foreach \m [count=\y] in {1, missing, 2}
	\node [every neuronv/.try, neuron \m/.try ] (embedding-\m) at (4,0.75-1.25*\y) {};
	
	\foreach \l [count=\i] in {1,2,3,4,5,p}
	\draw [<-] (input-\i) -- ++(-1,0)
	node [above, midway] {\Large $x_\l$};
		
		
	\foreach \m/\l [count=\y] in {1,2,3,4,5,missing,6}
	\node [every neuronb/.try, neuron \m/.try] (input_decode-\m) at (8,2.5-\y) {};
	
	\foreach \l [count=\i] in {1,\ell}
	\node [above] at (embedding-\i.north) {\Large $x^*_\l$};
	
	\foreach \l [count=\i] in {1,2,3,4,5,p}
	\draw [->] (input_decode-\i) -- ++(1,0)
	node [above, midway] {\Large $\widehat{x}_\l$};
	
	
	\foreach \i in {1,...,6}
	\foreach \j in {1,2}
	\draw [->, red] (input-\i) -- (embedding-\j);
	

	\foreach \i in {1,...,6}
	\foreach \j in {1, 2}
	\draw [->, blue] (embedding-\j) -- (input_decode-\i);
	
	
	\foreach \l [count=\x from 0] in {Input, Embedding, Reconstructed}
	\node [align=center, above] at (\x*4,2) {\l \\ layer};
	
	\end{tikzpicture}}

%% file: img/autoencoder.tex
\resizebox{\textwidth}{!}{\begin{tikzpicture}[x=1cm, y=1cm, >=stealth]
	
	
	\foreach \m/\l [count=\y] in {1,2,3,4, 5, missing,6}
	\node [every neuronr/.try, neuron \m/.try] (input-\m) at (0,2.5-\y) {};
	
	\foreach \m [count=\y] in {1,2,missing,3}
	\node [every neuronr/.try, neuron \m/.try ] (hidden-\m) at (2,1.5-\y*1.25) {};
	
	\foreach \m [count=\y] in {1, missing, 2}
	\node [every neuronv/.try, neuron \m/.try ] (embedding-\m) at (4,0.75-1.25*\y) {};
	
	\foreach \l [count=\i] in {1,2,3,4,5,p}
	\draw [<-] (input-\i) -- ++(-1,0)
	node [above, midway] {\Large $x_\l$};
	
	\foreach \l [count=\i] in {1,2,m}
	\node [above] at (hidden-\i.north) {\Large $h^{e}_\l$};

	
	\foreach \m [count=\y] in {1,2,missing,3}
	\node [every neuronb/.try, neuron \m/.try ] (hidden_decode-\m) at (6,1.5-\y*1.25) {};
	
	\foreach \m/\l [count=\y] in {1,2,3,4,5,missing,6}
	\node [every neuronb/.try, neuron \m/.try] (input_decode-\m) at (8,2.5-\y) {};
	
	\foreach \l [count=\i] in {1,\ell}
	\node [above] at (embedding-\i.north) {\Large $x^*_\l$};
	
	\foreach \l [count=\i] in {1,2,m}
	\node [above] at (hidden_decode-\i.north) {\Large $h^{d}_\l$};
	
	\foreach \l [count=\i] in {1,2,3,4,5,p}
	\draw [->] (input_decode-\i) -- ++(1,0)
	node [above, midway] {\Large $\widehat{x}_\l$};
	
	
	\foreach \i in {1,...,6}
	\foreach \j in {1,...,3}
	\draw [->, red] (input-\i) -- (hidden-\j);
	
	\foreach \i in {1,...,3}
	\foreach \j in {1, 2}
	\draw [->, red] (hidden-\i) -- (embedding-\j);

	\foreach \i in {1,...,3}
	\foreach \j in {1, 2}
	\draw [->, blue] (embedding-\j) -- (hidden_decode-\i);
	
	\foreach \i in {1,...,6}
	\foreach \j in {1,...,3}
	\draw [->, blue] (hidden_decode-\j) -- (input_decode-\i);
	
	\foreach \l [count=\x from 0] in {Input, Hidden \\ Encoder, Embedding, Hidden \\ Decoder, Reconstructed}
	\node [align=center, above] at (\x*2,2) {\l \\ layer};
	
	\end{tikzpicture}}

%% file: textual.tex

\section{Textual representations}\label{sec:nlp}

Our first example deals with textual representations. The field of computer science studying text is called natural language processing (NLP). A comprehensive introduction to NLP is given in \cite{jurafsky2000speech}. Individual words are categorical variables, usually represented in one-hot encodings. Texts (sentences, paragraphs or documents) are a series of words, so they have a categorical format and sequence type. This combination of types and formats make textual documents challenging to store in a typical design matrix. In this section, we present sources of documents for P\&C insurance. Then, we expand on the numerical analysis of words, including word embeddings. Then, we explain methods to extract representations of documents using word embeddings. 

\subsection{Source of documents for P\&C actuarial science}

A non-exhaustive list for sources of documents in P\&C actuarial science includes insurance contracts, insurance endorsements (amendments) modifying basic insurance contracts, claims adjuster notes, underwriter comments and email exchanges between the insurance company's agents with its clients. More textual information could be available to commercial insurance actuaries by scraping website information and customer reviews. Most textual applications of text in P\&C insurance research focus on claims adjuster notes, including classification of injury severity and type \cite{tixier2016automated}, classification of peril type from descriptions \cite{lee2020actuarial}, identifying large losses from descriptions \cite{baillargeon2021mining} \cite{lee2020actuarial}, \cite{sabban2020automatic}, and identifying fraudulent claims \cite{wang2018leveraging}. 

\subsection{Words}

Following \cite{jurafsky2000speech}, a dataset containing textual documents is called a corpus. A distinct word in a corpus is called a word type. The set of word types is called the vocabulary $V$, and its size is noted by $|V|$. A word in a corpus is called a token, and the number of tokens in a corpus is noted $N$. 

For example, consider a text composed of the word types \{\texttt{fire}, \texttt{home}, \texttt{cancer}, \texttt{flood}, \texttt{car}\}, so $|V| = 5$. We may use these features in machine learning models through one-hot encoding. 

As the corpus size increases, its size $|V|$ will also increase (the size is typically over 10 000), making these vectors very sparse. Additionally, we cannot perform mathematical operations on the vectors to determine correlations or similarities between words. The distributional hypothesis \cite{harris1954distributional} established the link between a word's context and its meaning. A better representation of a word would enable computational comparisons between similar words. For instance, we expect perils like \{\texttt{fire}, \texttt{flood}\} to be more similar to each other than the types of insurance \{\texttt{car}, \texttt{home}\} or to life insurance preconditions like \{\texttt{cancer}\}. 


\subsection{Word embeddings}

Word embeddings accommodate the distributional hypothesis by assigning a dense vector to each word. This vector contains continuous values, and its dimension is much smaller than $|V|$. A useful representation of a vocabulary in two dimensions would cluster perils and types of coverages separately. Returning to Figure \ref{fig:2dim}, we observe clusters of perils \{\texttt{Cat 1, Cat 4}\}, types of insurance \{\texttt{Cat 2}, \texttt{Cat 5}\} and an unrelated word \{\texttt{Cat 3}\}. Useful word embeddings trained on a large corpus would also produce a vector for a new coverage \texttt{liability} as a 6\textsuperscript{th} class close to the peril cluster.

Creating these representation vectors by hand (feature engineering) would be a tedious task. Instead, we use representation learning tools to extract useful representations from the data automatically. The typical method to model sequential variables is to construct a model using past data as input and predict the next observation in the sequence, i.e. a function of the shape $x_{T+1} = f(x_t, t = 1, \dots, T)$. To construct word embeddings, following \cite{collobert2008unified}, we also look at the future context to gather meaning, creating a function of the shape $x_t = f(x_{t-2}, x_{t-1}, x_{t+1}, x_{t+2})$. 

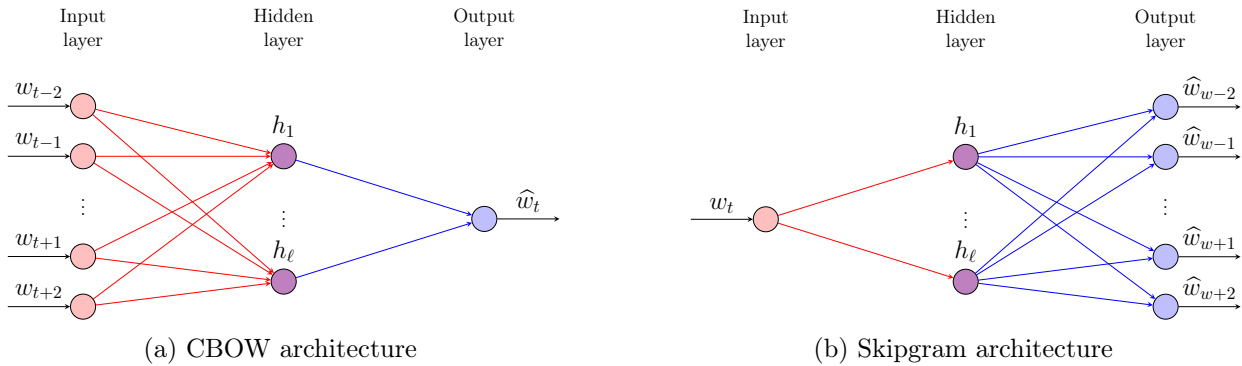
\begin{figure}[ht]
	\centering
	\begin{subfigure}[b]{0.45\textwidth}
		\centering
		\input{img/cbow.tex}
		
		\setlength{\belowcaptionskip}{-5pt} 
		\setlength{\abovecaptionskip}{-8pt} 
		\caption{CBOW architecture}
		
		\label{fig:cbow}
	\end{subfigure}
	\hfill
	\begin{subfigure}[b]{0.45\textwidth}
		\centering
		\input{img/skipgram.tex}		
		\setlength{\belowcaptionskip}{-5pt} 
		\setlength{\abovecaptionskip}{-8pt} 
		\caption{Skipgram architecture}
		\label{fig:skipgram}
	\end{subfigure}

	\setlength{\belowcaptionskip}{-5pt} 
	\caption{word2vec embedding models}
	\label{fig:word2vec}
\end{figure}

The first widespread word embedding models were word2vec, a pair of models presented in \cite{mikolov2013efficient, mikolov2013distributed} and illustrated in Figure \ref{fig:word2vec}. The diagram in Figure \ref{fig:cbow} is the continuous bag-of-word (CBOW) model, where the context predicts a word. The diagram in Figure \ref{fig:skipgram} presents the skipgram model, where a word predicts the context. The inputs and outputs of the model are one-hot encodings of individual words. Unlike the representation models presented in Section \ref{sec:representation}, the embeddings in word2vec are extracted from the parameter weight matrix (represented here by arrows). In the examples of Figure \ref{fig:word2vec}, there are two context words before and two after. The size of the context window will determine the type of word embeddings that are generated. Typical word embeddings have a dimension $\ell$ between 100 and 300, much shorter than $|V|$.

Word2vec is a classical approach to construct word embeddings, but we limit the scope of this paper to these models. For further explanations of word embeddings in actuarial science, see \cite{lee2020actuarial}. Training word embeddings is an active area of NLP research, the interested reader can look at \cite{devlin2018bert} and variants for modern word-embedding models using transformers.

\subsection{Document representations}

In NLP, word embeddings have great success as they compress knowledge from many NLP tasks in dense and compact vectors usable in modeling tasks \cite{collobert2011natural}. Today, most applications of NLP rely on pre-trained embeddings \cite{mikolov2017advances}. In this section, we will explain how to use these word embeddings to create document representations. 

Let $d$ be a document, consisting of $|d|$ word tokens. We will use the notation $w^{\langle i\rangle}$ to represent the $i$\textsuperscript{th} word in the document, $i = 1, \dots, |d|$. In addition, let $w^{\langle i\rangle *}$ be the word embedding of word $w^{\langle i\rangle}, i = 1, \dots, |d|$. We can summarize a document by calculating the document centroid, the average embedding vector in a document, given by
$$\boldsymbol{d}^* = \frac{1}{|d|}\sum_{i = 1}^{|d|} w^{\langle i\rangle *}.$$ 
Document centroids are useful to compare how similar two documents are, by using the cosine distance \eqref{eq:cosine} between two documents, $\text{cosine}(\boldsymbol{d}_1^*, \boldsymbol{d}_2^*)$. Applications of document similarity includes information retrieval, plagiarism detection and news recommender systems \cite{jurafsky2000speech}. 

\begin{figure}[ht]
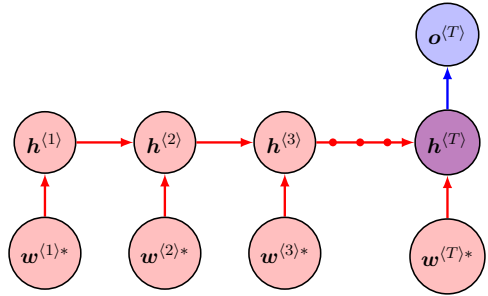

	\centering
	\include{img/rnn-one-col}
	\setlength{\belowcaptionskip}{-5pt} 
	\setlength{\abovecaptionskip}{-8pt} 
	\caption{Many-to-one recurrent neural network to predict the occurrence of a claim. Input: pre-trained word embeddings, output: estimated probability.}\label{fig:many-to-one}
\end{figure}

We can also create document representations for specific classification or regression tasks using the methods from Section \ref{sec:representation}. For example, consider an insurance company that writes many unique amendments to their primary insurance contracts. Calculating the amendments' effect on the frequency of claims is not as simple as with vectorial data since we cannot group amendments. In this case, we could train a model to predict a claim's occurrence, given the endorsement text. Since there is a sequence of word inputs and a single output, we can use a \textit{many-to-one} recurrent neural network as presented in Figure \ref{fig:many-to-one}.

To price a future insurance contract, we can use the output $\rnnvect{o}{T}$ from Figure \ref{fig:many-to-one} as an additive or multiplicative effect on the claim frequency (end-to-end learning), or extract an embedding from the hidden state $\rnnvect{h}{T}$ and use it as features in the ratemaking model (representation learning).  

%% file: img/cbow.tex
\resizebox{\textwidth}{!}{
	\begin{tikzpicture}[x=1cm, y=1cm, >=stealth]
	
	
	\foreach \m/\l [count=\y] in {1, 2, missing, 3, 4}
	\node [every neuronr/.try, neuron \m/.try] (input-\m) at (0,1.5-\y) {};
	
	\foreach \m [count=\y] in {1, missing, 2}
	\node [every neuronv/.try, neuron \m/.try ] (embedding-\m) at (4,0.75-1.25*\y) {};
	
	\foreach \l [count=\i] in {t-2, t-1, t+1, t+2}
	\draw [<-] (input-\i) -- ++(-1.5,0)
	node [above, midway] {\Large $w_{\l}$};
		
		
	\node [every neuronb/.try, neuron 1/.try] (input_decode-1) at (8,-1.75) {};
	
	\foreach \l [count=\i] in {1,\ell}
	\node [above] at (embedding-\i.north) {\Large $h_\l$};
	
	\foreach \l [count=\i] in {1}
	\draw [->] (input_decode-\i) -- ++(1.5,0)
	node [above, midway] {\Large $\widehat{w}_t$};
	
	
	\foreach \i in {1,...,4}
	\foreach \j in {1,2}
	\draw [->, red] (input-\i) -- (embedding-\j);
	


	\foreach \j in {1, 2}
	\draw [->, blue] (embedding-\j) -- (input_decode-1);

	\foreach \l [count=\x from 0] in {Input, Hidden, Output}
	\node [align=center, above] at (\x*4,1.5) {\l \\ layer};
	
, 	\end{tikzpicture}}

%% file: img/skipgram.tex
\resizebox{\textwidth}{!}{
	\begin{tikzpicture}[x=1cm, y=1cm, >=stealth]
	
	
	\node [every neuronr/.try, neuron 1/.try] (input-1) at (0,-1.75) {};
	
	\foreach \m [count=\y] in {1, missing, 2}
	\node [every neuronv/.try, neuron \m/.try ] (embedding-\m) at (4,0.75-1.25*\y) {};
	
	\foreach \l [count=\i] in {1}
	\draw [<-] (input-\i) -- ++(-1.5,0)
	node [above, midway] {\Large $w_t$};
		
		
	\foreach \m/\l [count=\y] in {1,2,missing,3,4}
	\node [every neuronb/.try, neuron \m/.try] (input_decode-\m) at (8,1.5-\y) {};
	
	\foreach \l [count=\i] in {1,\ell}
	\node [above] at (embedding-\i.north) {\Large $h_\l$};
	
	\foreach \l [count=\i] in {w-2, w-1, w+1,w+2}
	\draw [->] (input_decode-\i) -- ++(1.5,0)
	node [above, midway] {\Large $\widehat{w}_{\l}$};
	
	
	\foreach \j in {1,2}
	\draw [->, red] (input-1) -- (embedding-\j);
	

	\foreach \i in {1,...,4}
	\foreach \j in {1, 2}
	\draw [->, blue] (embedding-\j) -- (input_decode-\i);
	
	
	\foreach \l [count=\x from 0] in {Input, Hidden, Output}
	\node [align=center, above] at (\x*4,1.5) {\l \\ layer};
	
	\end{tikzpicture}}

%% file: img/rnn-one-col.tex
\resizebox{!}{0.17\textheight}{
	\begin{tikzpicture}[item/.style={circle,draw,thick,align=center},
	itemc/.style={item,on chain,join}]
	\node[item, fill=red!25] (A1) {$\rnnvect{h}{1}$};
	\node[item, right of = A1, xshift=3em, fill=red!25] (A2) {$\rnnvect{h}{2}$};
	\node[item, right of = A2, xshift=3em, fill=red!25] (A3) {$\rnnvect{h}{3}$};
	\node[item, right of = A3, xshift=5em, fill = violet!50] (AT) {$\rnnvect{h}{T}$};
	\draw[very thick,-latex, red] (A1) -- (A2);
	\draw[very thick,-latex, red] (A2) -- (A3);
	\draw[very thick,-latex, red] (A3) -- (AT);
	
	\foreach \X in {1,2,3,T} 
	{\draw[red, very thick,latex-] (A\X.south) -- ++ (0,-2em) node[below,item,black,fill=red!25] (x\X) {$\boldsymbol{w}^{\langle \X \rangle *}$};}
	\path (A3) -- (AT) node[midway,scale=2.5,font=\bfseries, red] {\dots};
	\draw[very thick,-latex, blue] (AT.north) -- ++ (0,2em) node[above,item,black, fill=blue!25] (hT) {$\rnnvect{o}{T}$};
	\end{tikzpicture}
}

%% file: visual.tex
\section{Visual representations}\label{sec:visual}

This section covers visual representations, the embeddings we create from image datasets. The field of computer science studying images is called computer vision (CV). The most imminent use of images in the ratemaking process includes extracting information to simplify the ratemaking process. For instance, an image of a house could provide information on roofing type or indicator variables like a garage's presence. While a human can perform this task (manual identification is the approach in \cite{kita2019google}) it can also be automated by training an image classification model. If this model performs well, there may be enough information within the image to capture other risk sources. We could then consider using the image as inputs to the ratemaking model. This section will provide details on sources of images and how to create representations with these sources. 

\subsection{Source of images for P\&C actuarial science}

A ratemaking model could use any image with a relationship with the insured product or the customer. Due to the potential discrimination of using a customer's image, we will restrict our analysis to images of the insured product. 

For homeowners insurance, using a house's image as input to the regression model could be beneficial. There are many methods to obtain the images, including asking the insured to provide a house picture. However, this may be cumbersome and discourage the customer from proceeding with the quoting process, so we seek an automatic method to extract this image from the internet. The two most readily available sources are aerial images (orthorectified images from satellites or planes) available through a provider like Google Satellite or facade images available from Google Streetview. These images are simple to obtain through a web API (Application Programming Interface). For instance, to download a Google Streetview image, we call the web API with the following command:
\begin{center}
	\small \texttt{https://maps.googleapis.com/maps/api/streetview?size=600x600\&location=LOCATION\&key=KEY}
\end{center}
where \texttt{LOCATION} is an address, and \texttt{KEY} is an API key provided by Google. Therefore, when a potential customer provides his or her address during the quoting process, the ratemaking model can automatically perform a call to this API, get the image and use it to provide the quote.

For car insurance, we could consider using the image of a car as input to the model. However, structured vehicle characteristics are more readily available than home information, so images may be unnecessary. 

\subsection{Data preprocessing}

\begin{figure}[ht]
	\centering
	\includegraphics[width=0.75\textwidth]{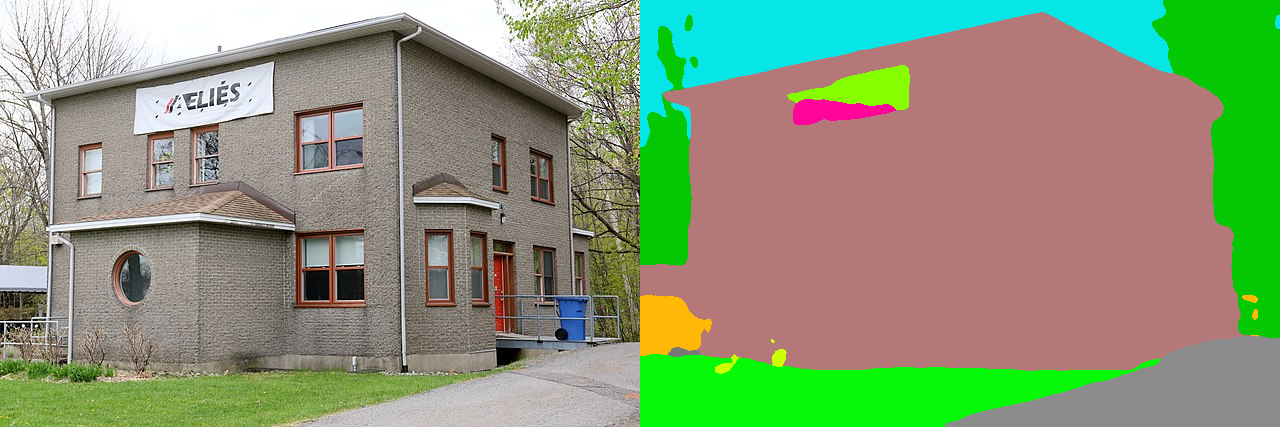}
	
	\setlength{\belowcaptionskip}{-5pt} 
	\caption{Image and semantic segmentation. Brown: building (59.82\%). Dark green: Tree (11.73\%). Blue: sky (9.92\%). Light green: grass (9.02\%). Gray: road (6.34\%). Other instances: <5\%. Original image: Simon Pierre Barrette, \textit{Marie-Sirois house}, 25 May 2019, CC BY-SA 4.0.}
	\label{fig:semantic-segmentation}
\end{figure}

Two main preprocessing steps to use images in our context includes identifying occlusions (by vehicles or trees) and centering the dwelling. To help with this step, we can use semantic segmentation techniques to identify instances in an image. For instance, applying \cite{zhou2017scene, zhou2018semantic}. Figure \ref{fig:semantic-segmentation} presents an image and its semantic instances using \texttt{ResNet50dilated + PPM\_deepsup}. 

%

The predicted instances can either be used to filter images (discard images where \texttt{house} or \texttt{building} instance is below threshold), center an image, or included as a fourth channel in the input image and let the representation model determine how to treat different instances. 

\subsection{Representations with convolutional autoencoders}

\begin{figure}[ht]
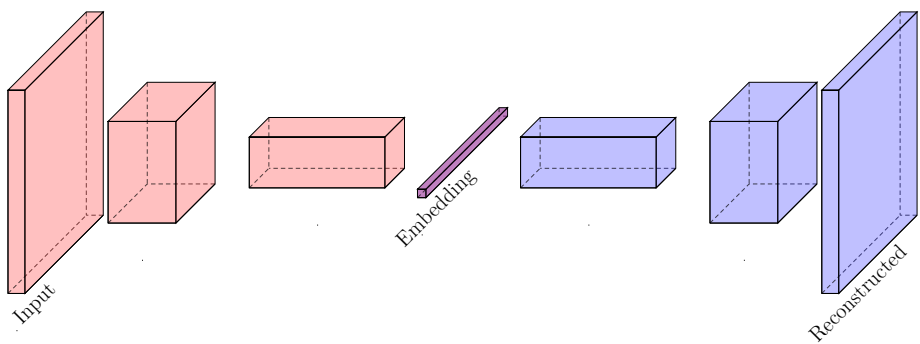

	\begin{center}
		\include{conv-ae}
	\end{center}
	\caption{Convolutional autoencoder to extract image embeddings}\label{fig:conv-ae}
\end{figure}

The first method to create representations of images is convolutional autoencoders, as presented in Figure \ref{fig:conv-ae}. Like the autoencoders that we presented in Section \ref{sec:representation}, the input and the output of the model are the images. In this model, the encoder is composed of convolutional operations. The final feature map is unrolled into a vector and followed by fully-connected layers. The model then rolls back the representation into a grid, followed by deconvolution steps and max unpooling, see \cite{dumoulin2016guide} for arithmetic details of these operations. 

\subsection{Representations with transfer learning}

Another method to create image representations is transfer learning. Consider a CNN trained to classify the content of an image on a large dataset like ImageNet. Then, the layer right before the prediction step will correspond to a semantic representation of an image's content. We can extract an embedding from this layer (i.e. discarding the final layer of the CNN).



%% file: conv-ae.tex
\resizebox{0.35\textwidth}{!}{
	\begin{tikzpicture}
	\tikzstyle{connection}=[ultra thick,every node/.style={sloped,allow upside down},draw=\edgecolor,opacity=0.7]
	\tikzstyle{copyconnection}=[ultra thick,every node/.style={sloped,allow upside down},draw={rgb:blue,4;red,1;green,1;black,3},opacity=0.7]
	
	\path[use as bounding box] (5, -2.5) rectangle (15, 4);

	\pic[shift={(0,0,0)}] at (0,0,0) 
	{Box={
			name=conv1,
			caption= ,
			xlabel={{, }},
			zlabel=\Large Input,
			fill=red,
			opacity=0.25,
			height=24,
			width=2,
			depth=24
		}
	};
	
	\pic[shift={(1.5,0,0)}] at (conv1-east) 
	{Box={
			name=conv2,
			caption= ,
			xlabel={{, }},
			zlabel=,
			fill=red,
			opacity=0.25,
			height=12,
			width=8,
			depth=12
		}
	};
	
	\pic[shift={(1.5,0,0)}] at (conv2-east) 
	{Box={
			name=conv3,
			caption= ,
			xlabel={{, }},
			zlabel=,
			fill=red,
			opacity=0.25,
			height=6,
			width=16,
			depth=6
		}
	};
	
	\pic[shift={(1.5,0,0)}] at (conv3-east) 
	{Box={
			name=soft1,
			caption=,
			xlabel={{" ","dummy"}},
			zlabel=\Large Embedding,
			fill=violet,
			opacity=0.5,
			height=1,
			width=1,
			depth=25
		}
	};
	
	\pic[shift={(1.5,0,0)}] at (soft1-east) 
	{Box={
			name=conv4,
			caption= ,
			xlabel={{, }},
			zlabel=,
			fill=blue,
			opacity=0.25,
			height=6,
			width=16,
			depth=6
		}
	};
	
	\pic[shift={(1.5,0,0)}] at (conv4-east) 
	{Box={
			name=conv5,
			caption= ,
			xlabel={{, }},
			zlabel=,
			fill=blue,
			opacity=0.25,
			height=12,
			width=8,
			depth=12
		}
	};
	
	\pic[shift={(1.5,0,0)}] at (conv5-east) 
	{Box={
			name=conv6,
			caption= ,
			xlabel={{, }},
			zlabel=\Large Reconstructed,
			fill=blue,
			opacity=0.25,
			height=24,
			width=2,
			depth=24
		}
	};
	
	\end{tikzpicture}
}

%% file: spatial.tex
\section{Spatial representations}\label{sec:geodata}

Geographical information is useful in insurance since it helps contextualize risks. For weather-related risks such as flooding, geographical information is crucial since location is a primary factor in assessing claim frequency. Insurance companies might also want to limit the number of houses it insures on the same street since, in the event a flood occurs, it will need to pay for many claims simultaneously, putting it in a difficult financial situation. For socio-demographic risks such as driving, habits depend on where drivers live: they are less likely to have accidents in rural areas since they use rarely frequented roads. When they have accidents, they tend to generate higher claims since crashes are more severe than in cities.

Insurers with a high risk exposure volume in a territory may rely on historical loss data to predict losses, using their experience to smooth previous losses and estimate risk relativities based on the one-hot representation of base territories. Insurers with low exposures in a territory must use external spatial information to predict future losses. This creates two issues: 
\begin{enumerate}
	\item much spatial information may be needed to model spatial effects adequately, and
	\item spatial effect relativities must vary smoothly in space.
\end{enumerate}
Spatial representations of risks will address these issues. This section will provide the ideas behind spatial representations, followed by an overview of the convolutional regional autoencoder (CRAE) \cite{blier2020encoding}, which designs a representation of geographical information. This model is based on convolutional autoencoders with an architecture similar to Figure \ref{fig:conv-ae}. Using spatial representations, we can use vectorial ratemaking models instead of using spatial models, therefore not requiring historical experience in a territory to produce a prediction. 

\subsection{Source of geographic information for P\&C actuarial science}

Many sources of geographic information could improve P\&C ratemaking. Experts estimate that 80\% of data is geographic \cite{chang2008introduction}. While we can represent spatial data in many formats (see \cite{blierwong2018correction} for details in actuarial science), we limit our study to geo-localized features (point pattern). These include most emerging external data like weather, crime, traffic and census information. 

\subsection{Geographic embeddings}

Territories are categorical variables, typically stored as one-hot encodings. For instance, consider a set of five cities \{\texttt{Matane}, \texttt{Montréal}, \texttt{Kuujjuaq}, \texttt{Sept-Îles}, \texttt{Québec}\}. An observation from Montréal would have a vector $\boldsymbol{e}_2$. Montréal and Québec are large urban areas, with heterogeneous populations and abundant services. For this reason, we expect these territories to be similar. Sept-Îles and Matane are smaller remote population centers with homogeneous populations. On the other hand, Kuujjuaq is a northern village composed of different populations and fewer services. The one-hot representation does not capture this similarity. When projecting territories in two dimensions, a good representation would create clusters of similar territories as in Figure \ref{fig:2dim}.

Creating these representations by hand would be tedious. Spatial embeddings learn these representations from data automatically. This section presents a method to construct these embeddings using census data, although other approaches are possible, see \cite{saeidi2015lower, zhang2019unifying, du2019beyond, hui2020predicting}. An added advantage of using spatial embeddings is that the embedding values vary smoothly, neighboring territories will have similar embedding values. This is desirable for insurance ratemaking since we expect neighbors to exhibit similar risk levels. Spatial embeddings are particularly interesting for insurers with little or no data in a territory. This could be the case for regions with low populations, a territory the insurer wishes to increase its portfolio, or establishing a new line of business in a territory. 

Consider an insurer with a portfolio in the province of Québec but who wishes to expand to the province of Ontario. When using word embeddings, we may predict quantities of interest for regions in Ontario using a model trained in the province of Québec. Toronto (new category 6) is a large urban region with similar characteristics to Montréal (category 2). Thus, we expect the spatial effects that generate spatial risk to be similar in both regions (more similar than Toronto and Kuujjuaq, category 3, for instance). The traditional actuarial technique to model spatial effects would be to increase the one-hot encoding dimension by one. However, we have no historical loss data to estimate the regression parameter associated with this new dimension and no way to let the model determine that this new dimension is similar to Montréal. Dense representations would keep the embedding dimension the same and learn similar vectors for Montréal and Toronto in an unsupervised way by using external data. Since their embedding values are similar, statistical models using dense embeddings as inputs will extrapolate Montréal effects to the new region of Toronto. Returning to Figure \ref{fig:2dim}, Toronto could have an embedding value of (3, 4.5), placing it closest to Montréal but farthest from Kuujjuaq.

\subsection{Dense spatial representations of geo-located information}

\begin{figure}[ht]
	\minipage{0.45\textwidth}
	\includegraphics[width=\linewidth]{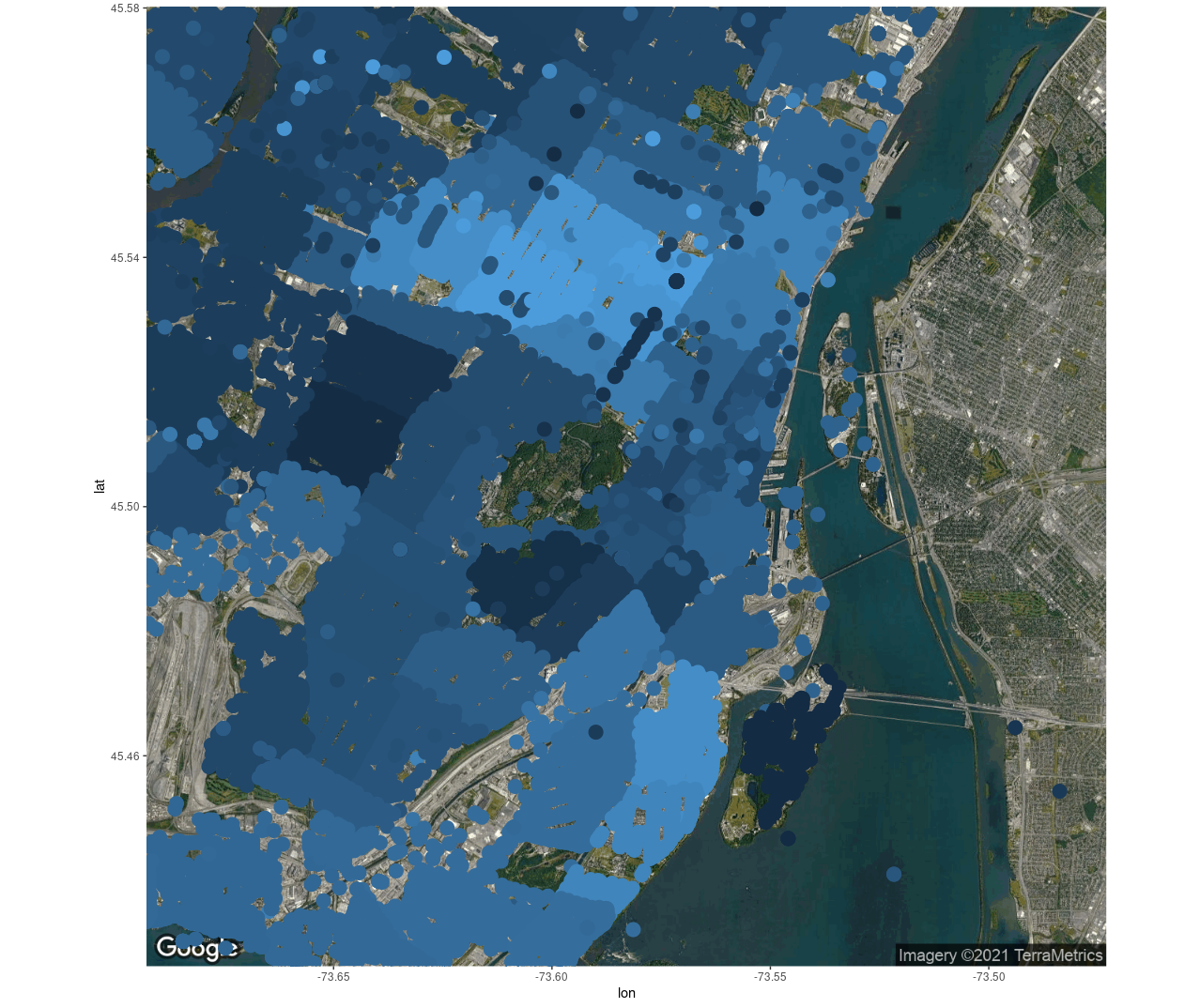}
	\caption{One dimensional embeddings}\label{fig:fsa}
	\endminipage\hfill
	\minipage{0.45\textwidth}
	\includegraphics[width=\linewidth]{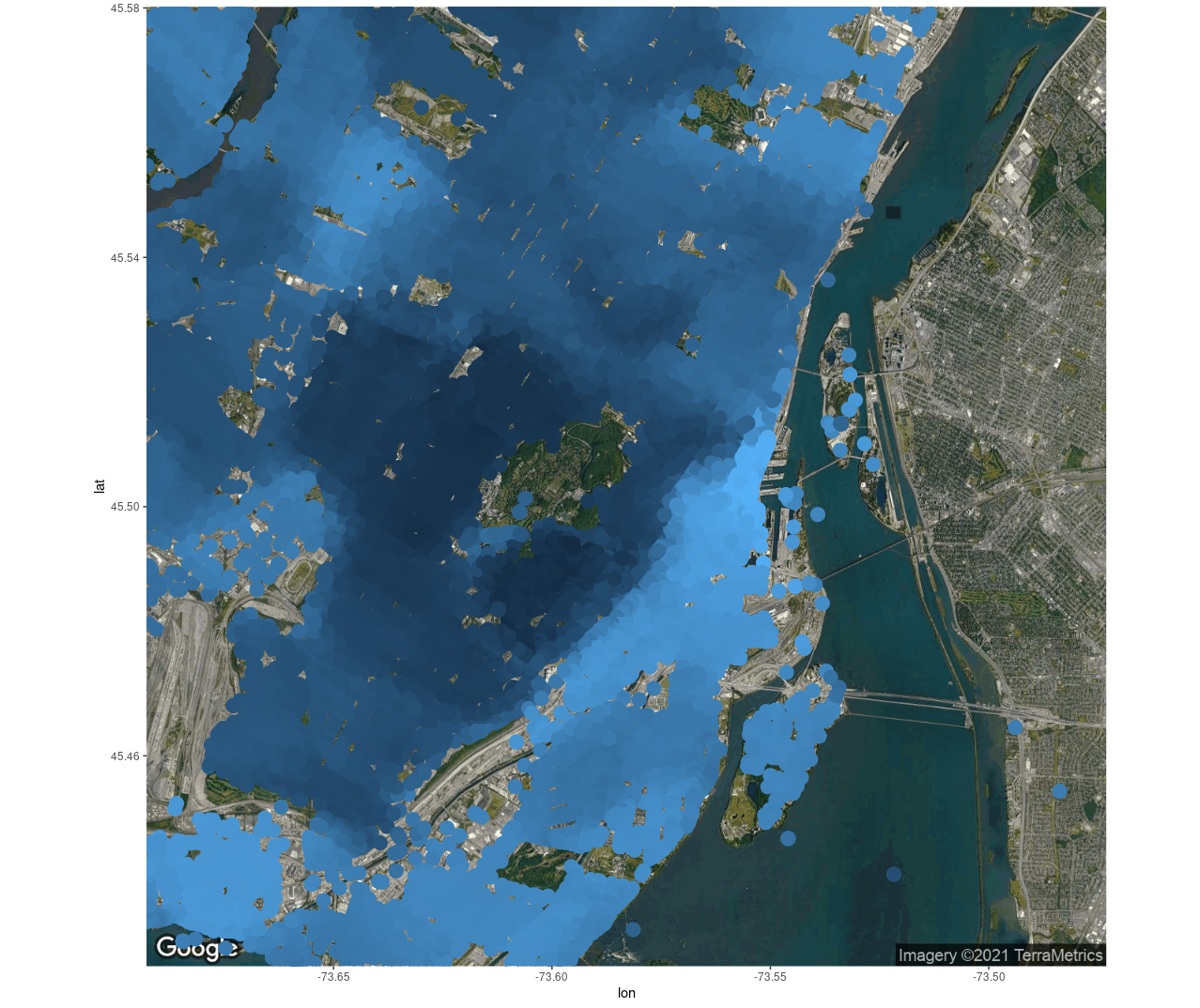}
	\caption{Two dimensional embeddings}\label{fig:fsaldu}
	\endminipage\hfill
	
\end{figure}

Spatial embeddings based on the one-hot encoding of territories cannot predict spatial effects in unobserved territories. Instead, we create dense representations based on external datasets with spatial information available across the entire territory of an insurer's portfolio. Consider a country performing census studies, providing summary statistics for territories across the country. We can use the information in this census as external data to construct the spatial representations. We can use the insured's address information to determine the summary information from citizens in the same geographic area. Suppose we have $p$ features in the census dataset. The vector of external geo-localized information for observation $i$ is noted $\boldsymbol{\gamma}_i \in \mathbb{R}^{p}$, consisting of every variable available in the census dataset. We can concatenate this vector with the original features for observation $i$ within a regression model. However, $p$ is usually large (over 2000 for Canada), so a dimension reduction procedure is required, either using PCA, autoencoders or transfer learning techniques. Figure \ref{fig:fsa} presents a map of a representation created using PCA on $\boldsymbol{\gamma}_i, i = 1, \dots, n$ to create an embedding vector $\boldsymbol{x}_i^*$. Each point corresponds to a postal code, and the color of the point is the intensity of one of the principal component values $\boldsymbol{x}_i^*$ at each location. 

We notice two problems from the plots of embeddings in Figure \ref{fig:fsa}. The first is that switching from one base territory to the next, there is a sudden jump in the embedding value. We expect spatial effects to vary smoothly in space. By using one-dimensional methods like PCA or autoencoders, neighboring information is not used to create the embeddings. These jumps are undesirable for insurance pricing since two neighboring customers could pay significantly different premiums if they each live on opposing sides of the border between two base territories. Another problem with the plot above is that geocoding is not reliable data. We can notice this problem in Figure \ref{fig:fsa} by observing individual points with embedding values different from those around it. A reason for this is that many datasets converting postal codes or addresses into GPS coordinates are crowd-sourced and not validated by central authorities. Another reason is that postal codes may not stay constant. Some may be created, merged or moved, all changing the embedding values or geocoding. Two-dimensional representation models based on GPS coordinates will circumvent these issues. 

\subsection{Convolutional regional autoencoder}

One implementation of dense geographic representations is the convolutional regional autoencoder (CRAE), introduced in \cite{blier2020encoding}. The main idea of CRAE is to transform spatial point pattern data into grid data by spanning a grid around a central location. This process creates an image with many channels that can be modeled using the same techniques as in Section \ref{sec:visual}. Suppose again that we have external data $\boldsymbol{\gamma}_i \in \mathbb{R}^{p}, i = 1, \dots, n$. Our objective remains to create an embedding vector $\boldsymbol{x}_i^*\in \mathbb{R}^{\ell}$ for $i = 1, \dots, n$. Instead of creating a representation model $f: \mathbb{R}^{p} \to \mathbb{R}^{\ell}$, we will use external data from the surrounding location as well. 

\begin{figure}[ht]
	\minipage{0.45\textwidth}
	\includegraphics[width=\linewidth]{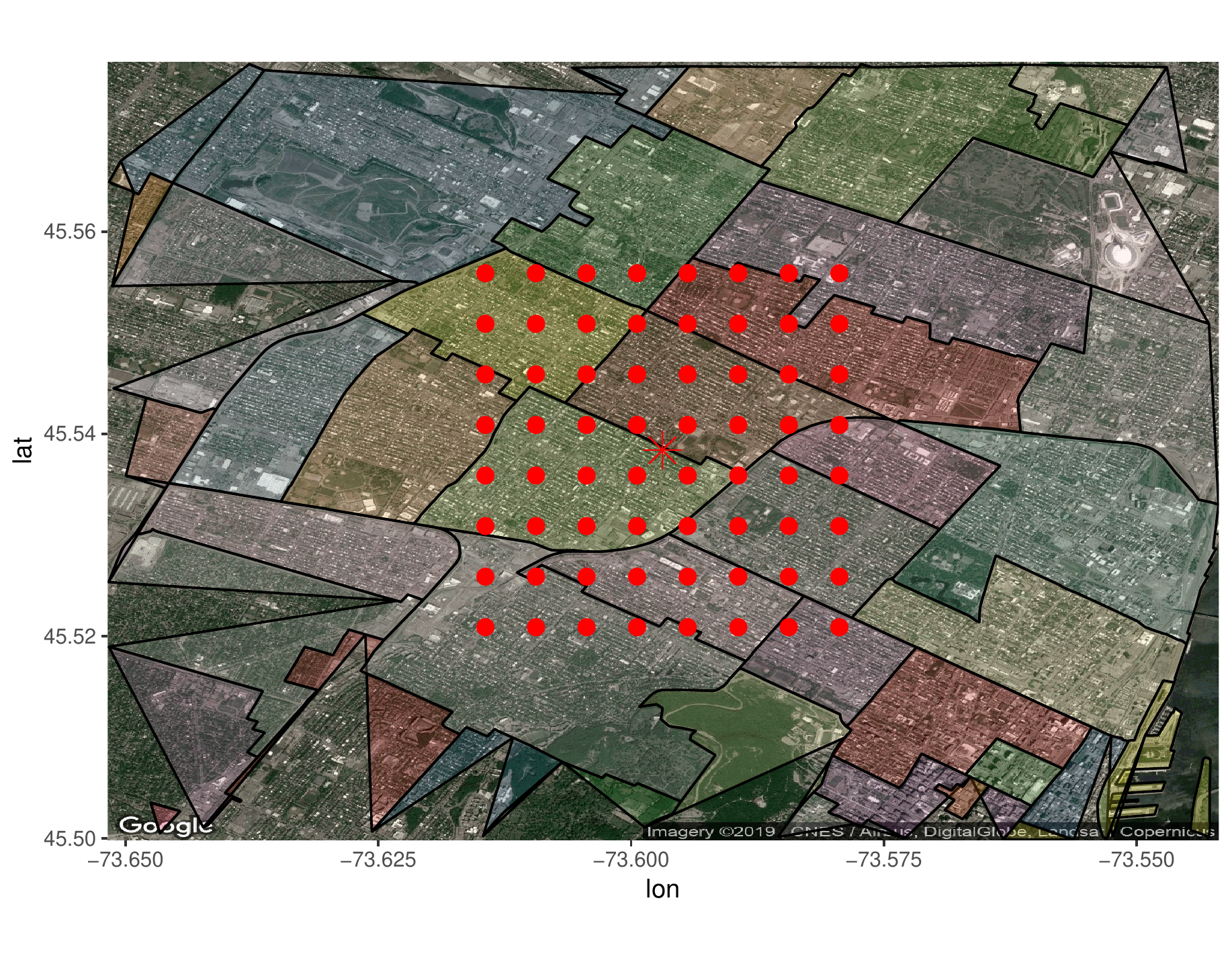}
	
	\setlength{\belowcaptionskip}{-5pt} 
	\setlength{\abovecaptionskip}{-8pt} 
	\caption{Grid around point}\label{fig:span}
	\endminipage\hfill
	\minipage{0.45\textwidth}
	\include{img/cube}
	
	\setlength{\belowcaptionskip}{-5pt} 
	\setlength{\abovecaptionskip}{-8pt} 
	\caption{Data square cuboid}\label{fig:grid}
	\endminipage\hfill
	
	\setlength{\belowcaptionskip}{-5pt} 
\end{figure}

To create a representation of the spatial information for the location of risk $i$ (noted by a star in Figure \ref{fig:span}), we use information from $\nu_i$, a set of geographic coordinates for observations around the location of risk $i$. A straightforward method to generate the set $\nu_i$ is to create a $q \times q$ grid around the location of risk $i, i = 1, \dots, n$ and the size of the set will be $|\nu_i| = q^2$. Since every element of the neighbor set belongs to a base territory (represented by polygons in Figure \ref{fig:span}), we can extract spatial information $\gamma_j, \forall~j \in \nu_{i}$, constructing a geographic data square cuboid, see Figure \ref{fig:grid}. 

The data square cuboid will contain census information (represented by the depth $p$ of the data square cuboid) for each neighbor in $\nu_i$ (represented by the grid). This new data representation can be interpreted the same way as an image, but the number of channels will be $p$ instead of 1 (grayscale) or 3 (color). The model presented in \cite{blier2020encoding} uses a convolutional autoencoder as presented in Figure \ref{fig:conv-ae}, and the embeddings are extracted using the encoder $f: \mathbb{R}^{q\times q\times p} \to \ell.$

CRAE has shown useful as predictors for downstream regression tasks. In Figure \ref{fig:fsaldu}, we intrinsically evaluate the quality of representations by plotting the values for one dimension of $\boldsymbol{x}^*$ generated by CRAE. The values are much smoother than Figure \ref{fig:fsa}, and there is no spatial discontinuity. The mislocated points are gone since our embedding method uses coordinates and not base territories. 


%% file: img/cube.tex
\resizebox{0.8\linewidth}{0.7\linewidth}{\begin{tikzpicture}
	
	\draw[thick, step=1cm] (-4,-4) grid (4,4);
	\tstar{0.05}{0.5}{8}{0}{thick,fill=red}
		\foreach \n in {-3.5, -2.5, ..., 3.5}
	\foreach \j in {-3.5, -2.5, ..., 3.5}
	\node at (\n, \j)[circle,fill,inner sep=3pt, color = red]{};
	
	\draw[thick] (-4, 4) -- (-12,12);
	\draw[thick] (-4, -4) -- (-12,4);
	\draw[thick] (-12,12) -- (-12,4);
	\draw[thick] (-12,12) -- (-4,12);
	\draw[thick] (-4,12) -- (4, 4);
	
	\draw[thick] (-4.5,4.5) -- (3.5, 4.5);
	\draw[thick] (-4.5,4.5) -- (-4.5, -3.5);	
	
	\foreach \n in {0.5, 1, 1.5, 7, 7.5}	
	\draw[thick] (-4 - \n, 4 + \n) -- (4 - \n, 4 + \n);
	\foreach \n in {0.5, 1, 1.5, 7, 7.5}	
	\draw[thick] (-4 - \n,4 + \n) -- (-4-\n, -4 + \n);	
	
	\foreach \n in {1, 2, 3}	
	\node at (0 - \n * 0.5, 3.75 + \n * 0.5) {Feature \n};
	
	\node at (-11.25 + 3.5, 11.75) {Feature $d$};
	\node at (-11.25 + 3.5 + 0.5, 11.75 - 0.5) {Feature $d$-1};
	
	\draw[loosely dotted, very thick] (-4.5,8.5) -- (-4.2,8.2);
	\draw[loosely dotted, very thick] (-8.5,4.5) -- (-8.2,4.2);
	
	\end{tikzpicture}}

%% file: conclusion.tex
\section{Conclusion}\label{sec:conclusion}

In this paper, we presented a unified framework for P\&C ratemaking in actuarial science with multisource data. To accomplish this, we split the P\&C ratemaking process into two steps: an encoder to create representations (the so-called representation model) and a regression model to perform actuarial tasks. A good encoder will create useful representations for the insurance process, such that the GLM regression model can be simple. We explained many advantages of this approach, including staying within the GLM framework (and retaining statistical properties of GLMs) but remaining as flexible as modern machine learning models. 

This framework can accept vectorial, image and sequential variables. We presented special cases of the framework for textual documents, images and geographic data. In forthcoming papers, we apply our representation framework in an actuarial pricing context. 

While most machine learning applications to ratemaking focused on improving end-to-end predictive models, our paper focuses on data. By creating useful representations of data, we can ultimately improve the performance of predictive models and retain the statistical properties of GLMs. While this approach is not novel in computer science, we believe this framework will help actuaries leverage better insights from data.

\section*{Acknowledgments}

The first author gratefully acknowledges support through fellowships from the Natural Sciences and Engineering Research Council of Canada (NSERC) and the Chaire en actuariat de l’Université Laval. This research was funded by NSERC (Cossette: 04273, Marceau: 05605), and the Chaire en actuariat de l’Université Laval (FO502320). We thank Cyril Blanc for discussions on the use of images in actuarial science, especially for the semantic segmentation step. We also thank Professor Thierry Duchesne for helpful comments. 